\newif\ifcol
\coltrue

\newif\ifbw
\bwfalse

\newif\ifdraft
\draftfalse
\newcommand{\pagesize}{letter}

\ifdraft
  \colfalse
\fi

\newlength{\plotwidth}
\ifdraft
\fi

\ifdraft
  \documentclass[11pt,draftcls,\pagesize,onecolumn,twoside]{IEEEtran}
\else
  \documentclass[10pt,journal,\pagesize,twoside]{IEEEtran}
\fi

\usepackage{url}

\usepackage{ifpdf}

\ifpdf
  \usepackage[pdftex]{hyperref}
\else
  \usepackage[ps2pdf]{hyperref}
\fi

\usepackage{booktabs}

\usepackage{cite}      % Written by Donald Arseneau
\ifpdf
  \usepackage[pdftex]{graphicx}
\else
  \usepackage[dvips]{graphicx}
\fi

\usepackage{subfigure} % Written by Steven Douglas Cochran

\usepackage{amssymb}    % Required for some math symbols uses, e.g. R for real numbers.

\usepackage{amsmath}    % From the American Mathematical Society

\newcommand{\eqn}[1]{(#1)}

\newcommand{\tbl}[1]{Table~#1}

\newcommand{\fig}[1]{Figure~#1}

\newcommand{\sectn}[1]{Section~#1}

\newcommand{\appn}[1]{Appendix~#1}

\newcommand{\thmref}[1]{Theorem~#1}

\newcommand{\eg}{\mbox{\it e.g.}}
\newcommand{\ie}{\mbox{\it i.e.}}

\newcommand{\arcmin}{\ensuremath{{}^\prime}}
\newcommand{\degrees}{\ensuremath{{^\circ}}}

\newcommand{\cmb}{{CMB}}
\newcommand{\cmbtext}{{cosmic microwave background}}

\newcommand{\wmap}{{WMAP}}
\newcommand{\wmaptext}{{Wilkinson Microwave Anisotropy Probe}}

\newcommand{\mexhat}{Mexican hat}

\newcommand{\healpix}{{\tt HEALPix}}

\newcommand{\stwofil}{{\tt S2FIL}}
\newcommand{\fastcswt}{{\tt FastCSWT}}
\newcommand{\comb}{{\tt COMB}}

\newcommand{\fwhm}{{FWHM}}
\newcommand{\snr}{{\rm SNR}}

\newcommand{\mf}{{MF}}
\newcommand{\saf}{{SAF}}

\newcommand{\spcend}{\ensuremath{\:}}
\newcommand{\img}{\ensuremath{{\rm i}}}
\newcommand{\cconj}{\ensuremath{\ast}}

\newcommand{\realsnz}{\ensuremath{\mathbb{R}^{+}_{\ast}}}
\newcommand{\integers}{\ensuremath{\mathbb{Z}}}

\newcommand{\ltwo}{\ensuremath{\mathrm{L}^2}}
\newcommand{\sphere}{\ensuremath{{\mathbb{S}^2}}}
\newcommand{\sothree}{\ensuremath{{\mathrm{SO}(3)}}}

\newcommand{\zerovect}{\ensuremath{\mathbf{0}}}

\newcommand{\zreal}{{\ensuremath{\rm{Re}}}}
\newcommand{\zimag}{{\ensuremath{\rm{Im}}}}

\newcommand{\opnexpv}{\ensuremath{E \left[}}
\newcommand{\clsexpv}{\ensuremath{\right]}}

\newcommand{\dx}{\ensuremath{\mathrm{\,d}}}
\newcommand{\dmu}[1]{\ensuremath{\dx \Omega(#1)}}

\newcommand{\deul}[1]{\ensuremath{\dx \varrho(#1)}}

\newtheorem{lemma}{Lemma}%[section]
\newtheorem{result}{Theorem}
\newtheorem{definition}{Definition}
\newcommand{\thmend}{{\mbox{}  \hfill \raggedright \ensuremath{\QEDopen}}}

\newcommand{\sa}{\ensuremath{\omega}}
\newcommand{\saa}{\ensuremath{\theta}}
\newcommand{\sab}{\ensuremath{\phi}}
\newcommand{\sas}{\ensuremath{\saa, \sab}}
\newcommand{\eul}{\ensuremath{\mathbf{\rho}}}
\newcommand{\euls}{\ensuremath{\eula, \eulb, \eulc}}
\newcommand{\eula}{\ensuremath{\alpha}}
\newcommand{\eulb}{\ensuremath{\beta}}
\newcommand{\eulc}{\ensuremath{\gamma}}
\newcommand{\el}{\ensuremath{\ell}}
\newcommand{\m}{\ensuremath{m}}
\newcommand{\n}{\ensuremath{{\m\p}}}
\newcommand{\elm}{\ensuremath{{\el\m}}}

\newcommand{\elmax}{\ensuremath{{\el_{\rm max}}}}
\newcommand{\p}{\ensuremath{^\prime}}

\newcommand{\scale}{\ensuremath{R}}
\newcommand{\scalenaut}{\ensuremath{R_0}}

\newcommand{\kron}[2]{\ensuremath{\delta_{{#1}{#2}}}}

\newcommand{\shfarg}[3]{\ensuremath{Y_{#1#2}({#3})}}
\newcommand{\shfargc}[3]{\ensuremath{Y_{#1#2}^\cconj({#3})}}

\newcommand{\shfc}[2]{\ensuremath{Y_{#1#2}^\cconj}}
\newcommand{\shc}[3]{\ensuremath{{#1}_{{#2}{#3}}}}
\newcommand{\shcc}[3]{\ensuremath{{#1}_{{#2}{#3}}^\cconj}}
\newcommand{\shcsp}[3]{\ensuremath{{#1}_{{#2},{#3}}}}

\newcommand{\aleg}[3]{\ensuremath{P_{#1}^{#2}({#3})}}

\newcommand{\jacobi}[4]{\ensuremath{P_{#1}^{(#2,#3)}({#4})}}

\newcommand{\dmatbig}{\ensuremath{D}}

\newcommand{\dmatsmall}{\ensuremath{d}}

\newcommand{\dlmnb}{\ensuremath{ \dmatsmall_{\m\n}^{\el}(\eulb) }}

\newcommand{\dil}{\ensuremath{\mathcal{D}}}
\newcommand{\dilp}{\ensuremath{\mathcal{D}_\pnorm}}

\newcommand{\rot}{\ensuremath{\mathcal{R}}}

\newcommand{\sky}{\ensuremath{f}}

\newcommand{\cocycle}{\ensuremath{\lambda}}

\newcommand{\sumlm}{\ensuremath{\sum_{\el=0}^{\infty} \sum_{\m=-\el}^\el}}
\newcommand{\sumlmb}{\ensuremath{\sum_{\el\m}}}

\newcommand{\fil}{\ensuremath{\Psi}}
\newcommand{\filcoeff}{\ensuremath{F}}
\newcommand{\lagrnmult}{\ensuremath{\mu}}
\newcommand{\lagrn}{\ensuremath{\mathcal{L}}}
\newcommand{\filvara}{\ensuremath{a}}
\newcommand{\filvarb}{\ensuremath{b}}
\newcommand{\filvarc}{\ensuremath{c}}
\newcommand{\filvardenom}{\ensuremath{\Delta}}
\newcommand{\pnorm}{\ensuremath{p}}
\newcommand{\pnormsep}{\ensuremath{,}}
\newcommand{\pnormtext}{\ensuremath{{\rm L}^\pnorm}-norm}
\newcommand{\scalepnorm}{\ensuremath{{\scale\pnormsep\pnorm}}}
\newcommand{\scalenautpnorm}{\ensuremath{{\scalenaut\pnormsep\pnorm}}}
\newcommand{\za}[2]{\ensuremath{\shc{a}{#1}{#2}}}
\newcommand{\zb}[2]{\ensuremath{\shc{b}{#1}{#2}}}
\newcommand{\zc}[2]{\ensuremath{\shc{c}{#1}{#2}}}
\newcommand{\zd}[2]{\ensuremath{\shc{d}{#1}{#2}}}
\newcommand{\zcsp}[2]{\ensuremath{\shcsp{c}{#1}{#2}}}
\newcommand{\zdsp}[2]{\ensuremath{\shcsp{d}{#1}{#2}}}
\newcommand{\detect}{\ensuremath{\Gamma}}
\newcommand{\source}{\ensuremath{s}}
\newcommand{\beam}{\ensuremath{b}}
\newcommand{\noise}{\ensuremath{n}}
\newcommand{\instnoise}{\ensuremath{r}}
\newcommand{\obsig}{\ensuremath{f}}

\newcommand{\amp}{\ensuremath{A}}
\newcommand{\tmpl}{\ensuremath{\tau}}
\newcommand{\noisecl}{\ensuremath{C}}
\newcommand{\amphat}{\ensuremath{\hat{A}}}
\newcommand{\eulshat}{\ensuremath{\hat{\alpha},\hat{\beta},\hat{\gamma}}}
\newcommand{\fconsta}{\ensuremath{A_{\el\pnorm}}}
\newcommand{\fconstb}{\ensuremath{B_{\el\m}}}
\newcommand{\fconstaz}{\ensuremath{A_{0\pnorm}}}
\newcommand{\fconstazo}{\ensuremath{A_{01}}}

\begin{document}
\title{Optimal filters on the sphere}

% 

\author{Jason~D.~McEwen,~Michael~P.~Hobson,~and~Anthony~N.~Lasenby
%        Michael~Shell,~\IEEEmembership{Member,~IEEE,}
%        John~Doe,~\IEEEmembership{Fellow,~OSA,}
%        and~Jane~Doe,~\IEEEmembership{Life~Fellow,~IEEE}% <-this % stops a space
\thanks{Manuscript received 26 January, 2007.}%  The work of J.~D.~McEwen is supported by a Commonwealth (Cambridge) Scholarship from the Association of Commonwealth Universities and the Cambridge Commonwealth Trust.}% <-this % stops a space
\thanks{The authors are with the 
        Astrophysics Group, Cavendish Laboratory, Cambridge, UK.}
\thanks{E-mail: mcewen@mrao.cam.ac.uk (J.~D.~McEwen)}}
% note the % following the last \IEEEmembership and also the first \thanks - 
% instance, "A\textbf{} \textbf{}B" will typeset as "A B" not "AB". If you want
% "AB" then you have to do: "A\textbf{}\textbf{}B"
\markboth{IEEE Transactions on Signal Processing,~Vol.~--, No.~--,~November~2006}%
{McEwen \MakeLowercase{\textit{et al.}}: Optimal filters on the sphere}
% *** Note that you probably will NOT want to include the author's name in ***
% *** the headers of peer review papers.                                   ***

% If you want to put a publisher's ID mark on the page
% (can leave text blank if you just want to see how the
% text height on the first page will be reduced by IEEE)
%\pubid{0000--0000/00\$00.00~\copyright~2002 IEEE}

%\specialpapernotice{(Invited Paper)}

\maketitle

%------------------------------------------------------------------------------

% Abstract.
%------------------------------------------------------------------------------
% Abstract
%------------------------------------------------------------------------------

\begin{abstract}
  We derive optimal filters on the sphere in the context of detecting
  compact objects embedded in a stochastic background process.  The
  matched filter and the scale adaptive filter are derived on the
  sphere in the most general setting, \mbox{allowing} for directional
  template profiles and filters.  The performance and relative merits
  of the two optimal filters are discussed.  The application of
  optimal filter theory on the sphere to the detection of compact
  objects is demonstrated on simulated data.  A naive detection
  strategy is adopted, with an initial aim of illustrating the
  application of the new filter theory derived on the sphere.
  Nevertheless, this simple object detection strategy is demonstrated
  to perform well, even at low signal-to-noise ratio.
\end{abstract}

% Keywords.
\begin{keywords}
Filtering, spheres, convolution.
\end{keywords}

% Will be ignored for other modes.
\IEEEpeerreviewmaketitle

% Main body.

%=============================================================================
\section{Introduction}
\label{sec:intro}
%=============================================================================

\PARstart{T}{he} detection of compact objects embedded in a stochastic background is a problem experienced in many branches of physics and signal processing, and as such it has received considerable attention.  Many of these applications are restricted to flat Euclidean space, such as the one-dimensional line or the two-dimensional plane.  However, data are often measured or defined on other manifolds, such as the two-sphere.  
For example, applications where data are defined on the sphere are found in 
planetary science, %(\eg\ \cite{wieczorek:2006,wieczorek:1998,turcotte:1981}), 
geophysics, %(\eg\ \cite{whaler:1994,swenson:2002,simons:2006}), 
computer vision, %(\eg\ \cite{ramamoorthi:2004}),
quantum chemistry and % (\eg\ \cite{choi:1999,ritchie:1999}) and
astrophysics. % (\eg\ \cite{bennett:1996,bennett:2003a}).
Astrophysical applications include observations made on the celestial sphere of the \cmbtext\ (\cmb) (\eg\ \cite{hinshaw:2006}), a relic radiation of the Big Bang.  \cmb\ observations may be contaminated by localised foreground emission due to point sources or Sunyaev-Zel'dovich (SZ) effects induced by the hot intergalactic gas bound to clusters of galaxies \cite{sz:1970}.  These foreground emissions must be separated from \cmb\ observations in order to provide cleaned \cmb\ data for cosmological analysis or simply so that they may be studied in their own right.  Furthermore, other primordial physical phenomena may imprint localised signatures in the \cmb\ that are of interest (\eg\ cosmic strings \cite{kaiser:1984}).
The extension of optimal filter theory to the sphere would allow compact objects embedded in full-sky \cmb\ data to be detected in an analogous manner to that performed in the plane currently, \ie\ by filtering the observed field to enhance the contribution of embedded objects relative to the stochastic background, before attempts are made to recover the embedded objects from the filtered field using various classification schemes.

When adopting the filtering approach to object detection a number of criteria may be imposed so that the filter kernels are in some sense optimal.  The optimal matched filter (\mf) has been used extensively in many branches of physics and signal processing, and in the context of detecting point sources and SZ emission in \cmb\ observations 
made on small patches of the sky, which are assumed to be approximately flat, by \cite{tegmark:1998} and \cite{haehnelt:1995} respectively.  Other optimal filters in Euclidean space, such as the scale adaptive filter (\saf), have been derived by \cite{sanz:2001,herranz:2002}.  It is shown by \cite{sanz:2001} that the \mexhat\ wavelet on the plane is in fact the optimal \saf\ for the special case of detecting a Gaussian source embedded in a white noise background.  
The \mexhat\ wavelet has been used to detect objects embedded in \cmb\ data on small patches of the sky \cite{cayon:2000}. %,vielva:2001}.  
Furthermore, both the \mf\ and \saf\ (including the \mexhat\ wavelet) have been applied to simulated time ordered \cmb\ data to detect point sources \cite{barreiro:2003}. %,lopez:2005c}.  
Some debate exists, however, over the advantage of the \saf\ relative to the usual \mf\ \cite{vio:2004}.

All of the works discussed previously are limited to Euclidean space.  To analyse full-sky \cmb\ observations the techniques described previously must be extended to a spherical manifold.  
As a consequence of a full-sky analysis, it should be noted that the statistical properties of the background process are assumed to be stationary over the sphere. 
The \mexhat\ wavelet analysis has been extended to the sphere and applied to point source detection in the \cmb\ by \cite{vielva:2003}.  The extension of optimal filter theory to the sphere has been derived recently by \cite{schaefer:2004} and applied to simulated data to detect the SZ effect \cite{schaefer:2006}.   However, the optimal filter theory derived by \cite{schaefer:2004} is restricted to azimuthally symmetric source profiles and filters on the sphere.  In this paper we re-derive optimal filter theory on the sphere, making the extension to the more general class of non-azimuthally symmetric source profiles and filters.  In addition, we generalise to \pnormtext\ preserving dilations in order to highlight some minor amendments to previous works.

The remainder of this paper is organised as follows. In \sectn{\ref{sec:prelim}} some mathematical preliminaries are presented before the object detection problem is formalised in \sectn{\ref{sec:problem}}.  Derivations of the \mf\ and \saf\ on the sphere are presented in \sectn{\ref{sec:filters}}.  In \sectn{\ref{sec:simulations}} the new optimal filter theory is applied to detect objects embedded in simulated data, in the \mf\ case.  Concluding remarks are made in \sectn{\ref{sec:conclusions}}.

%=============================================================================
\section{Mathematical Preliminaries}
\label{sec:prelim}
%=============================================================================

It is necessary to outline some mathematical preliminaries
before attempting to derive optimal filters on the sphere.  Firstly, harmonic analysis on the two-sphere \sphere\ and on the rotation group \sothree\ is reviewed.  By making all assumptions and definitions explicit we hope to avoid any confusion over the conventions adopted.  A dilation operator is then defined on the sphere, before we review filtering on the sphere.  
Dilation and filtering are described in both real and harmonic space.

\subsection{Harmonic representations}
\label{sec:prelim_harm}

We consider the space of square integrable functions $L^2(\sphere,\dmu{\sa})$ on the unit two-sphere \sphere, where \mbox{$\dmu{\sa} = \sin\saa \dx\saa \dx\sab$} is the usual rotation invariant measure on the sphere and $\sa = (\sas) \in \sphere$ denotes spherical coordinates with colatitude \saa\ and longitude \sab.
A square integrable function on the sphere $f \in L^2(\sphere,\dmu{\sa})$ may be represented by the spherical harmonic expansion
% 
% \begin{displaymath}
%\linebreak[4]
\mbox{$
f(\sa) = \sum_{\el=0}^\infty \sum_{\m=-\el}^\el \shc{f}{\el}{\m} \shfarg{\el}{\m}{\sa}
$},
% \spcend ,
% \end{displaymath}
where the spherical harmonic coefficients are given by the usual projection on to the spherical harmonic basis functions:
% \begin{displaymath}
\mbox{$
\shc{f}{\el}{\m} 
= 
\int_{\sphere}
f(\sa) \:
\shfargc{\el}{\m}{\sa}  \:
\dmu{\sa}
$}.
% \spcend .
% \end{displaymath}
The \cconj\ denotes complex conjugation.
We adopt the Condon-Shortley phase convention where the normalised 
spherical harmonics are defined by \cite{varshalovich:1989}
\begin{displaymath}
\shfarg{\el}{\m}{\sa} = (-1)^\m \sqrt{\frac{2\el+1}{4\pi} 
\frac{(\el-\m)!}{(\el+\m)!}} \: 
\aleg{\el}{\m}{\cos\saa} \:
{\rm e}^{\img \m \sab}
\spcend ,
\end{displaymath}
where $\aleg{\el}{\m}{x}$ are the associated Legendre functions.  Using this normalisation the orthogonality of the spherical harmonic functions is given by
\begin{equation}
\label{eqn:shortho}
\int_\sphere
\shfarg{\el}{\m}{\sa} \:
\shfargc{\el\p}{\m\p}{\sa} \:
\dmu{\sa}
=
\kron{\el}{\el\p} \kron{\m}{\m\p}
\spcend ,
\end{equation}
where $\delta_{ij}$ is the Kronecker delta function.  

To perform filtering on the sphere one must define translations on the sphere, which may be naturally represented by rotations.  Rotations on the sphere $\rot$ are characterised by the elements of the rotation group \sothree, which we parameterise in terms of the three Euler angles \mbox{$\rho=(\euls)\in \sothree$}, where $\eula\in[0,2\pi)$, $\eulb\in[0,\pi]$ and $\eulc\in[0,2\pi)$.\footnote{We adopt the $zyz$ Euler convention corresponding to the rotation of a physical body in a \emph{fixed} co-ordinate system about the $z$, $y$ and $z$ axes by \eulc, \eulb\ and \eula\ respectively.}  The rotation of $f$ is defined by
% \begin{displaymath}
$
[\rot(\eul) f](\sa) \equiv f(R_\eul^{-1} \sa) 
$,
% \spcend ,
% \end{displaymath}
where $R_\eul$ is the rotation matrix corresponding to \rot(\eul).  It is also useful to characterise the rotation of a function on the sphere in harmonic space.  
% The rotation of a spherical harmonic basis function may be represented by a sum of weighted harmonics of the same \el\ \cite{brink:1999,ritchie:1999}:
% \begin{equation}
% \label{eqn:shrot:1}
% \left[\rot(\eul)\shf{\el}{\m}\right](\sa) = 
% \sum_{\n=-\el}^{\el} 
% \dmatbig_{\n\m}^{\el}(\eul) \: \shfarg{\el}{\n}{\sa}
% \spcend ,
% \end{equation}
% where the Wigner functions $\dmatbig_{\m\n}^{\el}(\eul)$ are described below.  From \eqn{\ref{eqn:shrot:1}} it is trivial to show that 
The harmonic coefficients of a rotated function are related to the coefficients of the original function by 
\begin{equation}
\label{eqn:shrot:2}
\left[\rot(\eul) f \right]_\elm = 
\sum_{\n=-\el}^{\el} 
\dmatbig_{\m\n}^{\el}(\eul) \: 
f_{\el\n}
\spcend .
\end{equation}
% Note carefully the distinction between the indices of the Wigner functions in \eqn{\ref{eqn:shrot:1}} and \eqn{\ref{eqn:shrot:2}}.
% 
The Wigner functions may be decomposed as \cite{varshalovich:1989}%\cite{brink:1999,ritchie:1999}
\begin{equation}
\label{eqn:d_decomp}
\dmatbig_{\m\n}^{\el}(\euls)
= {\rm e}^{-\img \m\eula} \:
\dmatsmall_{\m\n}^\el(\eulb) \:
{\rm e}^{-\img \n\eulc}
\spcend ,
\end{equation}
where the real polar \dmatsmall-matrix is defined by \cite{varshalovich:1989}
\begin{align}
\label{eqn:wignerd_b}
  \dlmnb =&
  \sqrt{\frac{(\el+\n)! (\el-\n)!}{(\el+\m)! (\el-\m)!}}
  \left( \sin\frac{\eulb}{2} \right)^{\n-\m} \nonumber \\ & \times
  \left( \cos\frac{\eulb}{2} \right)^{\n+\m} 
  \jacobi{\el-\n}{\n-\m}{\n+\m}{\cos\eulb}
  \spcend ,
\end{align}
where $\jacobi{\el}{a}{b}{\cdot}$ are the Jacobi polynomials.
% For alternative more practical definitions of the \dmatsmall-matrices see \cite{mcewen:thesis} and references therein.
% Alternative more practical definitions of the \dmatsmall-matrices are given in \cite{kostelec:2003,mcewen:thesis}.
% \ifcol
%   \begin{eqnarray*}
%   \lefteqn{\dmatsmall^\el_{\m \n}(\eulb)  = 
%   \sum_{t = \max(0, \m - \n)}^{\min(\el + \m, \el - \n)} (-1)^t} \nonumber\\
%   && \times \frac{
%   \left[(\el+\m)! \, (\el-\m)! \, (\el+\n)! \, (\el-\n)! \, \right]^{1/2}}
%   {(\el+\m-t)! \, (\el-\n-t)! \, (t+\n-\m)! \, t!} \nonumber \\
%   && \times \left[\cos\!\left(\frac{\eulb}{2} \right)\right]^{2\el+\m-\n-2t}
%   \left[\sin \! \left(\frac{\eulb}{2} \right)\right]^{\n-\m+2t}
%   \spcend ,
%   \end{eqnarray*}
% \else
%   \begin{eqnarray*}
%   \dmatsmall^\el_{\m \n}(\eulb) & = & 
%   \sum_{t = \max(0, \m - \n)}^{\min(\el + \m, \el - \n)} (-1)^t
%   \frac{
%   \left[(\el+\m)! \, (\el-\m)! \, (\el+\n)! \, (\el-\n)! \, \right]^{1/2}}
%   {(\el+\m-t)! \, (\el-\n-t)! \, (t+\n-\m)! \, t!} \nonumber \\
%   && \times \left[\cos\!\left(\frac{\eulb}{2} \right)\right]^{2\el+\m-\n-2t}
%   \left[\sin \! \left(\frac{\eulb}{2} \right)\right]^{\n-\m+2t}
%   \spcend ,
%   \end{eqnarray*}
% \fi
% and the sum over $t$ is defined so that the arguments of the factorials are non-negative.  
The Wigner functions satisfy the orthogonality condition
\begin{equation}
\label{eqn:wigner_ortho}
\int_\sothree 
\dmatbig_{\m n}^{\el}(\eul) \:
\dmatbig_{\m\p n\p}^{\el\p\cconj}(\eul) \:
\deul{\eul} 
=
\frac{8\pi^2}{2\el+1} \:
\kron{\el}{\el\p} \kron{\m}{\m\p} \kron{n}{n\p}
\spcend ,
\end{equation}
where $\deul{\eul} = \sin\eulb \dx \eula  \dx \eulb  \dx \eulc$.
Recursion formulae are available to compute rapidly the Wigner  \dmatsmall-matrices (see \eg\ \cite{risbo:1996}).
% in the basis of either complex \cite{risbo:1996,choi:1999} or real \cite{ivanic:1996,blanco:1997} spherical harmonics.

%=============================================================================
\subsection{Dilation on the sphere}

\begin{figure}
\centerline{
  \includegraphics[width=86mm]{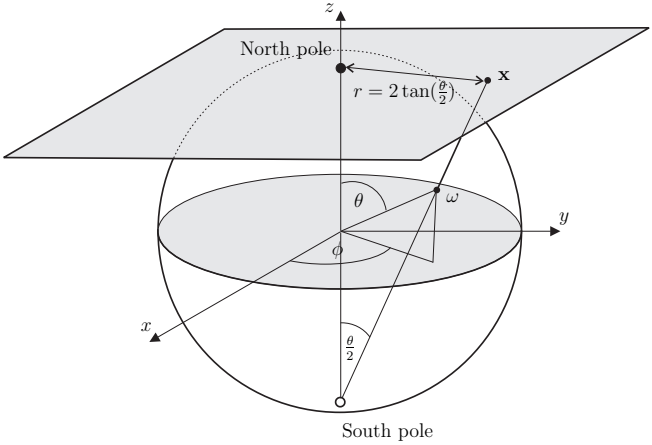}}
\caption[Stereographic projection]{Stereographic projection of the
  sphere on to the plane.}
\label{fig:stereographic_projection}
\end{figure}

To perform filtering on the sphere at various scales a spherical dilation operator must be defined.  We adopt the spherical dilation operator first defined by \cite{antoine:1999} %,antoine:1998}
to derive a continuous wavelet transform on the sphere.  In this case, the stereographic projection is used to project the sphere on to the plane (see \fig{\ref{fig:stereographic_projection}}).  
It is shown by \cite{wiaux:2005} that the stereographic projection operator is the unique unitary, radial and conformal diffeomorphism between the sphere and the plane.
Dilations on the sphere are defined by: (a) first lifting the sphere to the plane using the stereographic projection; (b) performing the usual Euclidean dilation in the plane; (c) reprojecting back on to the sphere using the inverse stereographic projection.  The spherical dilation operator derived by \cite{antoine:1999,wiaux:2005}
%antoine:1998,
preserves the \ltwo-norm of functions.  We generalise to \pnormtext\ preserving dilations in order to highlight the impact of this choice on the final optimal filter equations derived in \sectn{\ref{sec:filters}}.  Although the choice of \pnorm\ is not of considerable practical interest, different works have implicitly assumed different \pnorm\ values, hence the explicit dependence given here should allow a direct comparison with previous works.
% The scale of an \pnormtext\ preserving dilation is denoted by \scalepnorm.
Definitions of the dilation and inverse dilation (contraction) operators follow, accompanied by short proofs to show that these operators do indeed preserve the \pnormtext.

\begin{definition}
The \pnormtext\ preserving spherical dilation is defined by
% \ifcol
%   \begin{align}
%   f_\scalepnorm(\sas) &\equiv
%   [\dil(\scalepnorm) f](\sas) \nonumber \\ 
%   &\equiv
%   [\cocycle(\scale,\saa)]^{1/\pnorm} \: f(\saa_{1/\scale}, \phi)
%   \spcend ,
%   \end{align}
% \else
  \begin{equation}
  f_\scalepnorm(\sas) \equiv
  [\dilp(\scale) f](\sas) \equiv
  [\cocycle(\scale,\saa)]^{1/\pnorm} \: f(\saa_{1/\scale}, \phi)
  \spcend ,
  \end{equation}
% \fi
where scale $\scale \in \realsnz$, \pnormtext\ $\pnorm \in \integers^{+}_{\ast}$,
$\tan(\saa_\scale/2) = \scale \tan(\saa/2)$ and the
$\cocycle(\scale,\saa)$ cocycle term introduced to preserve the
appropriate norm is defined by 
\begin{equation}
\label{eqn:cocycle}
\cocycle(\scale,\saa) = \frac{4 \: \scale^2}
{[(\scale^2-1)\cos{\saa}+(\scale^2+1)]^2}
\spcend .
\end{equation}
\thmend
\end{definition}

\begin{proof}
See \appn{\ref{sec:appn_dilation}}.
\end{proof}

\noindent A contraction, or inverse spherical dilation, may be
similarly defined and follows trivially from Definition 1.
\begin{definition}
The \pnormtext\ preserving inverse spherical dilation is defined by
\begin{equation}
[\dilp^{-1}(\scale) f](\sas) \equiv
[ \cocycle(\scale, \saa_\scale) ] ^{-1/\pnorm} \:
f(\saa_\scale,\sab)
\spcend .
\end{equation}
%where $\tan(\saa_\scale/2) = \scale \tan(\saa/2)$.
% and the cocycle is defined by \eqn{\ref{eqn:cocycle}}.
\thmend
\end{definition}

% \begin{proof}
% The inverse \pnormtext\ preserving spherical dilation defined above it proved trivially by apply the % dilation and inverse to recover the original function:
% % has the
% % effect of transforming a dilated function back to the original
% % function, thus it is indeed the inverse:
% %   \begin{displaymath}
% $
%   [\dilp^{-1}(\scale) \:
%   \dilp(\scale) \: f](\sas) 
% %   = 
% %   \dil^{-1}(\scalepnorm)
% %   \left \{
% %   [\cocycle(\scale,\saa)]^{1/\pnorm} \:
% %   f(\saa_{1/\scale},\sab) \right \} \nonumber \\
% %   &= 
% %   [\cocycle(\scale,\saa_\scale)]^{-1/\pnorm} \:
% %   [\cocycle(\scale,\saa_\scale)]^{1/\pnorm} \:
% %   f(\sas) \nonumber \\
%   = f(\sas) \nonumber
% $.
% %   \spcend .
% %   \end{displaymath}
% \end{proof}

The spherical dilation and contraction operations are
performed about the north pole.  A dilation about any other point on
the sphere may be performed by rotating that point to the north pole,
dilating, then rotating back to the original position.  
% Filter kernels defined on the sphere are centred on the north
% pole usually, in which case the initial rotation is not required.

%=============================================================================
% Dilation in harmonic space

When constructing optimal filters on the sphere it is necessary to
consider the spherical harmonic coefficients of dilated functions.  We
derive an intermediate result here to express the harmonic
coefficients of the dilated function in terms of the original
undilated function and dilated spherical harmonic functions. 

\begin{lemma}
\label{lemma:sh_dil}
The harmonic coefficients of a dilated function may be given either
by projecting the dilated function on to each spherical harmonic, or
equivalently by projecting the original undilated function on to
contracted spherical harmonics, scaled by the
appropriate cocycle factor:
\begin{equation}\label{eqn:sh_dil}
\shc{[\dilp(\scale) f]}{\el}{m} 
= \int_\sphere
f(\sas) \:
\frac{\shfargc{\el}{m}{\saa_\scale,\sab}}
{[\cocycle(\scale,\saa_\scale)]^{1-1/\pnorm}} 
\dmu{\sas}
\spcend .
\end{equation}
Indeed, for the 2-norm case ($\pnorm=2$) one finds $1-1/\pnorm=1/\pnorm$
and the spherical harmonics are contracted in accordance with the
definition of the inverse dilation:
  \begin{displaymath}
  \shc{[\dil_2({\scale}) f]}{\el}{m} 
  = \int_\sphere
  f(\sas) \:
  [ \dil_2^{-1}({\scale})
  \shfc{\el}{\m}](\sas) \:
  \dmu{\sas}
  \spcend .
  \end{displaymath} \thmend
\end{lemma}

\begin{proof}
Consider the spherical harmonic coefficients of a dilated function
\begin{displaymath}
\shc{[\dilp(\scale) f]}{\el}{m} 
= \int_\sphere 
[\cocycle(\scale,\saa)]^{1/\pnorm} \:
f(\saa_{1/\scale},\sab) \:
\shfargc{\el}{m}{\sas} \:
\dmu{\sas}
\spcend .
\end{displaymath}
By performing a change of variables this may be represented by 
  \begin{displaymath} 
  [\dilp (\scale) f]_{\el \m}  =
  \int_\sphere 
  f(\saa,\sab) \:
  \frac{\shfargc{\el}{m}{\saa_\scale,\sab}}{[\cocycle(\scale,\saa_\scale)]^{-1/\pnorm}} \:
  \dmu{\saa_\scale, \sab}
  \spcend ,
  \end{displaymath}
from which \eqn{\ref{eqn:sh_dil}} follows by noting 
$\dmu{\saa_\scale, \sab} = [\cocycle(\scale,\saa_\scale)]^{-1} \dmu{\sas}$.
\end{proof}

%=============================================================================
\subsection{Filtering on the sphere}

Filtering on the sphere is performed analogously to filtering on the plane; it is the \emph{spherical} convolution of the filter kernel with the analysed signal.  The analogue of translations on the sphere are rotations, thus the filtered field is given by the directional spherical convolution
\begin{equation}
\label{eqn:filtering}
\filcoeff(\eul, \scalepnorm) =
\int_{\sphere}
\sky(\sa) \:
[\rot(\eul) \fil_\scalepnorm]^\cconj(\sa) \;
\dmu{\sa}
\spcend ,
\end{equation}
where $\fil\in L^2(\sphere,\dmu{\sa})$ is the filter kernel.
All orientations in the rotation group \sothree\ are considered, thus
directional structure is naturally incorporated.
The filter equation \eqn{\ref{eqn:filtering}} is identical to the analysis formula of the continuous wavelet transforms derived on the sphere by \cite{antoine:1999,wiaux:2005}, %,antoine:1998,sanz:2006,mcewen:2006:cswt2}, 
hence our fast algorithm to evaluate this equation \cite{mcewen:2006:fcswt} may be applied to compute the filtered field rapidly. % (an alternative fast algorithm is presented in \cite{wiaux:2005c}).
For an example of the directional spherical filtering operation applied to Earth data see \cite{mcewen:2006:fcswt}.

When deriving optimal filters on the sphere it is often convenient to represent the filtering operation in harmonic space.  
Representing the analysed function and the rotated filter kernel in harmonic space,
one may rewrite \eqn{\ref{eqn:filtering}} as
\begin{equation}
\filcoeff(\eul, \scalepnorm) =
\sum_{\el =0}^{\infty} \:
\sum_{m=-\el }^{\el } \:
\sum_{m\p=-\el }^{\el }
\shc{\sky}{\el}{m} \:
\dmatbig_{\m\m\p}^{\el\cconj}(\eul) \,
\shcc{(\fil_\scalepnorm)}{\el}{m\p}
\spcend .
\end{equation}
For a filter centred on the north pole %(\ie\ a filter that is not rotated)
the harmonic representation of the filtering operation reduces to
\begin{equation}
\filcoeff(\zerovect, \scalepnorm) =
\sumlmb
\shc{\sky}{\el}{m} \:
\shc{(\fil_\scalepnorm)}{\el}{m}^\cconj
\spcend ,
\end{equation}
where here and subsequently we use the shorthand notation $\sumlm \equiv \sumlmb$.
Furthermore, for the special case of a
filter kernel that is azimuthally symmetric, %(\ie\ invariant under azimuthal rotations when located at the north pole), 
the filter is dependent on $\saa$ only (and not \sab) and, consequently, the
filtered field is independent of the value of $\eulc$.  
% Filtering with an azimuthally
% symmetric kernel may be written as
% \begin{equation}
% \filcoeff(\eul, \scalepnorma, \eulb) =
% \int_{\sphere}
% \sky(\sa) \:
% [\rot(\eula,\eulb,0) \fil_\scalepnorm]^\cconj(\sa) \;
% \dmu{\sa}
% \spcend ,
% \end{equation}
% where 
In this case, the spherical harmonic coefficients of the filtered field for a
particular scale are simply given by
\begin{equation}
\shc{[\filcoeff(\scalepnorm)]}{\el}{m} =
\sqrt{\frac{4 \pi}{2\el+1}} \:
\shc{\sky}{\el}{m} \:
\shc{(\fil_\scalepnorm)}{\el}{0}^\cconj
\spcend .
\end{equation}

%=============================================================================
\section{Problem Formalisation}
\label{sec:problem}
%=============================================================================

In this section we formalise the problem of detecting compact objects
embedded in a stochastic background noise process, and propose optimal
filtering to enhance the detection of such objects.  The formulation given here is similar to that of \cite{schaefer:2004} but is considered in the most general sense, allowing asymmetric templates of
various amplitude, position and orientation.  Removing the assumption
of an azimuthally symmetric template, the case considered in
previous works \cite{schaefer:2004,schaefer:2006}, introduces a number of complications as
the spherical filtering operation may no longer be represented
in harmonic space simply as a product of spherical harmonic coefficients.  

%==============================================================================
\subsection{Formulation}

Consider an observed field on the sky $\obsig(\sa)$ consisting of
a number of sources $\source_i(\sa)$ embedded in a
stochastic background process $\noise(\sa)$.  
Observations are likely to be made in the presence of additional instrument noise $\instnoise(\sa)$ also.
The observed field is obtained by measuring (convolving) the actual
field with some beam $\beam(\sa)$. 
% and may be expressed as
% % \begin{equation}
% $
% \obsig(\sa) = \beam(\sa)
% \conv
% \Bigl (
% \sum_i \source_i(\sa) + \noise(\sa)
% \Bigr )
% + \instnoise(\sa)
% $,
% % \spcend ,
% % \end{equation}
% where $\conv$ denotes spherical convolution with an azimuthally symmetric kernel (\cf\ \eqn{\ref{eqn:filtering}} with $\eulc=0$).  
The beam response and instrumental noise may be absorbed into the template and the statistical properties of the background, therefore these components may be included trivially when required.
Without loss of generality, the observed field may therefore be represented by
\begin{equation}
\obsig(\sa) = 
\sum_i \source_i(\sa) + \noise(\sa)
\spcend .
\end{equation}
% Essentially, we absorb the beam and receiver noise into the template and the statistical properties of the background, therefore these components may be included trivially when required.
% 
Each source may be represented in terms of its amplitude
$\amp_i$ and source profile
$\source_i(\sa) = \amp_i \: \tmpl_i(\sa)$,
where $\tmpl_i(\sa)$ is a dilated and rotated version of 
the source profile $\tmpl(\sa)$ of default dilation centred
on the north pole, \ie\
\mbox{$\tmpl_i(\sa) = \rot(\eul_i) \dil(\scale_i\pnormsep\pnorm)  \:
\tmpl(\sa)$}. 
One wishes to recover the parameters $\{\amp_i,\scale_i,\eul_i\}$ that describe each source amplitude, scale and position/orientation respectively.
The stochastic background process is assumed to be a zero-mean 
\mbox{$\opnexpv\noise(\sa)\clsexpv = 0$},
homogeneous and isotropic Gaussian random field, 
fully characterised by the spectrum
\begin{equation}
\opnexpv\shc{\noise}{\el}{\m}
\shc{\noise}{\el\p}{\m\p}^\cconj\clsexpv
= \kron{\el}{\el\p}
\kron{\m}{\m\p} \:
\noisecl_\el
\spcend ,
\end{equation}
where $\opnexpv \cdot \clsexpv$ denotes the expectation operator.
To facilitate the detection of compact sources, the observed
field is filtered using \eqn{\ref{eqn:filtering}} to enhance the source contribution relative to the background noise process.  The choice of filter kernels that are in some sense optimal is addressed next.

%==============================================================================
\subsection{Filter constraints}
\label{sec:filter_constraints}

Various optimal filters may be defined by imposing different
constraints on the filtered field.
Without loss of generality, we derive optimal filters for the
detection of a single source located at the north pole (hence we drop the $i$ subscript that denotes source index).  Sources located at
other positions and orientations are found by rotating the optimal filter, \ie\ by considering filtered field coefficients over \sothree.  

The following filter characteristics may be imposed when
constructing optimal filters:
\renewcommand{\labelenumi}{(\roman{enumi})}
\begin{enumerate}

\item {Unbiased:} The filtered field is an unbiased estimator of the
  source amplitude at the source position, \ie\ 
  $\opnexpv \filcoeff(\zerovect,\scalepnorm) \clsexpv=\amp$.

\item {Minimum variance:} The filtered field has minimum variance at the
  source position.%, \ie\ minimise $\sigma_\filcoeff^2(\zerovect,\scalepnorm)$.
% , where the variance is given by 
%   $\sigma_\filcoeff^2(\zerovect,\scalepnorm)=
%   \opnexpv | \filcoeff(\zerovect,\scalepnorm) | ^2 \clsexpv - 
%   | \opnexpv \filcoeff(\zerovect,\scalepnorm) \clsexpv |^2$.

\item {Local extremum in scale:} The expected value of the filtered
field has an extremum with respect to scale, \ie\
  $\frac{\partial}{\partial\scale} 
  \opnexpv\filcoeff(\zerovect,\scalepnorm)\clsexpv =0$ at $\scale=\scalenaut$.

\end{enumerate}
Imposing criteria (i) and (ii) only one obtains the usual 
\mf.  Imposing the additional constraint (iii)
one obtains the \saf, first
introduced on the plane by \cite{sanz:2001}.  This additional constraint imposes an extremum at the unknown scale $\scalenaut$.  In the derivation of the \saf\ this unknown scale drops out and the final filter expressions are independent of \scalenaut.  The additional constraint imposed when constructing the \saf\ provides a means to estimate
unknown source scales by checking for a maximum in scale but at
a cost of reduced gain.  
This issue is discussed in more detail in \sectn{\ref{sec:filters_comparison}}.

%==============================================================================
\subsection{Filtered field statistics}
\label{sec:field_stats}

In order to derive optimal filters on the sphere it is necessary to determine first 
expressions for the mean and variance of the filtered field at the
source position.
The filtered field mean at the source position is given by
\begin{equation}
\opnexpv \filcoeff(\zerovect,\scalepnorm) \clsexpv
= \amp \: 
\sumlmb
\shc{\tmpl}{\el}{m} \:
\shc{(\fil_\scalepnorm)}{\el}{m}^\cconj
\spcend .
\end{equation}
The filtered field variance at the source position is given by
\begin{equation}
\sigma_\filcoeff^2(\zerovect,\scalepnorm) =
\opnexpv | \filcoeff(\zerovect,\scalepnorm)|^2 \clsexpv - 
| \opnexpv \filcoeff(\zerovect,\scalepnorm) \clsexpv |^2
\spcend .
\end{equation}
The first term becomes
  \begin{displaymath}
  \opnexpv |\filcoeff(\zerovect,\scalepnorm)|^2 \clsexpv
  = 
  |\opnexpv \filcoeff(\zerovect,\scalepnorm) \clsexpv |^2
  +
  \sumlmb
  \noisecl_\el  \:
  \left | \shc{(\fil_\scalepnorm)}{\el}{m} \right|^2 \nonumber
  \spcend ,
  \end{displaymath}
where we have relied on the fact that the stochastic noise process has zero mean and is homogeneous and isotropic.  The variance is therefore given by
\begin{equation}
\label{eqn:fil_variance}
\sigma_\filcoeff^2(\zerovect,\scalepnorm)
=
\sumlmb \:
\noisecl_\el  \:
\left | \shc{(\fil_\scalepnorm)}{\el}{m} \right |^2
\spcend .
\end{equation}
The filtered field variance at the source position is also used to determine the error on amplitude estimates. %made using optimal filters.

%=============================================================================
\section{Optimal Filters}
\label{sec:filters}
%=============================================================================

In this section we derive the \mf\ and \saf\ on the sphere for an arbitrary template profile.  
The extension of optimal filter theory to the sphere has been derived already by \cite{schaefer:2004} for the special case of azimuthally symmetric source profiles.  
We re-derive optimal filter theory on the sphere here, making the extension to the more general class of non-azimuthally symmetric source profiles and filters.  In addition, we generalise to \pnormtext\ preserving dilations in order to highlight some minor amendments to previous works.
We show that the resultant optimal filters reduce to the expected definitions for azimuthally symmetric template profiles and also that, in the flat, continuous limit, these forms reduce to the optimal filter definitions derived on the plane.  To conclude this section we compare the performance of the \mf\ and \saf\ on the sphere and discuss the relative merits of each filter.

%=============================================================================
\subsection{Matched filter}
\label{sec:mf}

% The optimal \mf\ on the sphere is defined by the following theorem.
  \begin{result}
  \label{thm:mf}
  The optimal \mf\ defined on the sphere is obtained by
  imposing criteria (i) and
  (ii) defined in \sectn{\ref{sec:filter_constraints}}, \ie\ by solving the
  constrained optimisation  problem:
  \begin{displaymath}
  \min_{{\rm w.r.t.}\: \shc{(\fil_\scalepnorm)}{\el}{m} }
  {\sigma_\filcoeff^2(\zerovect,\scalepnorm)} \:\:\:\:
  \end{displaymath}
such that
  \begin{equation}
  \label{eqn:mf_constraint}
  \opnexpv \filcoeff(\zerovect,\scalepnorm) \clsexpv=\amp
  \spcend .
  \end{equation}
  The spherical harmonic coefficients of the resultant \mf\
  are given by
%   \boxedeqn{
  \begin{equation}
  \label{eqn:mf}
%  \boxed{
  \shc{(\fil_\scalepnorm)}{\el}{m}=
  \frac{\shc{\tmpl}{\el}{m}}
  {
  \filvara \:
  \noisecl_\el
  }
%  }  
  \spcend ,
% }  
\end{equation}
  where
%   \boxedeqn{
  \begin{equation}
  \label{eqn:filvara}
% \boxed{
  \filvara =
  \sumlmb
  \noisecl_\el^{-1} |\shc{\tmpl}{\el}{m}|^2
% }
  \spcend .
% }
  \end{equation}
  \thmend
  \end{result}

% -- MF Proof -------------------------
\begin{proof}
See \appn{\ref{sec:appn_mf}}.
\end{proof}

A measure of the capability of an optimal filter to detect an embedded
source is given by the \emph{maximum detection level}, defined by
\begin{equation}
\detect \equiv \frac{\opnexpv \filcoeff(\zerovect,\scalepnorm) \clsexpv}
{\sigma_{\filcoeff,{\rm min}}(\zerovect,\scalepnorm)}
\spcend .
\end{equation}
The minimum variance of the filtered field is found by substituting the expression for the \mf\ given by \eqn{\ref{eqn:mf}} into \eqn{\ref{eqn:fil_variance}}.  One finds that the \mf\ filtered field minimum variance is given by 
$\sigma_{\filcoeff,{\rm min}}^2(\zerovect,\scalepnorm)=\filvara^{-1}$, thus 
the maximum detection level for the \mf\ is
\begin{equation}
\detect_{\rm \mf}
=
\filvara^{1/2} \: \amp
\spcend .
\end{equation}

%=============================================================================
\subsection{Scale adaptive filter}
\label{sec:saf}

% The derivation of the \saf\ presented here is based on
% the approach first discussed by \cite{schaefer:2004} in the context of azimuthally symmetric templates and filters on the sphere.  We make the extension here to the more general case of asymmetric templates and filters on the sphere and make some other minor amendments.

To construct the optimal \saf\ defined on the sphere
we first recast criterion (iii) in a form expressed in
terms of the filter spherical harmonic coefficients and \emph{not}
the coefficients of the differentiated filter.  A suitable form is given
by the following lemma.

\begin{lemma}
\label{lemma:criterion_iii}
Optimal filter criterion (iii), imposing a local extremum in scale,
may be recast in the following more applicable form:
\begin{equation}
\label{eqn:constraint_iii}
\sumlmb \:
\shc{(\fil_\scalenautpnorm)}{\el}{\m}^\cconj
(
\fconsta
\shc{\tmpl}{\el}{\m}
-
\fconstb
\shcsp{\tmpl}{\el-1}{m}
) = 0
\spcend ,
\end{equation}
where 
% \begin{equation}
$\fconsta \equiv \el + 2/\pnorm - 1$
% \end{equation}
and
% \begin{equation}
$\fconstb \equiv (\el^2 - m^2)^{1/2}$.
% \spcend .
% \end{equation}
\thmend
\end{lemma}
The proof of this lemma may be found in \cite{mcewen:thesis} (the proof is not repeated here since it requires multiple pages).
% for a proof of this lemma.
% This lemma is proved in \appn{\ref{sec:appn}}.
It is now possible to derive the following theorem, the
definition of the optimal \saf\ defined on the sphere.

% -- SAF Result -------------------------
  \begin{result}
  \label{thm:saf}
  The optimal \saf\ defined on the sphere is obtained by imposing criteria
  (i), (ii) and (iii) defined in  \sectn{\ref{sec:filter_constraints}}, \ie\
  by solving the constrained optimisation problem:
  \begin{displaymath}
  \min_{{\rm w.r.t.}\: \shc{(\fil_\scalenautpnorm)}{\el}{m}}
  {\sigma_\filcoeff^2(\zerovect,\scalepnorm)}
  \end{displaymath}
  such that
  \begin{equation}
  \label{eqn:saf_constraint_i}
  \opnexpv \filcoeff(\zerovect,\scalepnorm) \clsexpv=\amp 
  \end{equation}
  and
  \begin{equation}
  \label{eqn:saf_constraint_ii}
  \left.
  \frac{\partial}{\partial\scale} \opnexpv\filcoeff(\zerovect,\scalepnorm)\clsexpv\right|_{\scale=\scalenaut}=0
  \spcend .
  \end{equation}
  The spherical harmonic coefficients of the resultant \saf\
  are given by
    \begin{equation}
    \shc{(\fil_\scalenautpnorm)}{\el}{m} =    
    \frac{ \filvarc \shc{\tmpl}{\el}{m} -
    \filvarb (
    \fconsta
    \shc{\tmpl}{\el}{m} 
    -
    \fconstb
    \shcsp{\tmpl}{\el-1}{m}
    )
    }{\filvardenom \noisecl_\el}
    \spcend ,
    \label{eqn:saf}
    \end{equation}
  where
  \begin{equation}
  \filvarb = \sumlmb \noisecl_\el^{-1} \shc{\tmpl}{\el}{m}
  ( 
  \fconsta
  \shc{\tmpl}{\el}{m}^\cconj
   -
  \fconstb
  \shcsp{\tmpl}{\el-1}{m}^\cconj
  )
  \spcend , 
  \label{eqn:filvarb}
  \end{equation}
  \begin{equation}
  \filvarc = \sumlmb \noisecl_\el^{-1} \bigl | 
  \fconsta
  \shc{\tmpl}{\el}{m} 
  -
  \fconstb
  \shcsp{\tmpl}{\el-1}{m}
  \bigr |^2
  \spcend ,
  \label{eqn:filvarc}
  \end{equation}
  \begin{equation}
  \label{eqn:filvardenom}
  \filvardenom = \filvara \filvarc - |\filvarb|^2
  \end{equation}
  and $\filvara$ is defined by \eqn{\ref{eqn:filvara}}. \thmend
  \end{result}

% -- SAF Proof -------------------------
\begin{proof}
See \appn{\ref{sec:appn_saf}}.
\end{proof}

The minimum variance of the filtered field is found by substituting the expression for the \saf\ given by \eqn{\ref{eqn:saf}} into \eqn{\ref{eqn:fil_variance}}.  One finds that the \saf\ filtered field minimum variance is given by 
$\sigma_{\filcoeff,{\rm min}}^2(\scalenautpnorm,\zerovect)=\filvarc\filvardenom^{-1}$,
thus the maximum detection level for the \saf\ is
\begin{equation}
\detect_{\rm \saf}
=
\filvarc^{-1/2}
\filvardenom^{1/2} \amp
\spcend .
\end{equation}

%=============================================================================
\subsection{Azimuthally symmetric case}

In this subsection we show that the expressions derived above for arbitrary template profiles reduce to the forms expected for azimuthally symmetric templates (the case considered by \cite{schaefer:2004}).
The spherical harmonic coefficients of azimuthally \mbox{symme\-tric} functions on the sphere are non-zero for $\m=0$ only, hence the general filter expressions that we derive for a directional template should reduce to the symmetric result when setting $\m=0$.  The definition of the filter variables used herein differ slightly to those defined by \cite{schaefer:2004} (which we denote by $\filvara\p$, $\filvarb\p$, $\filvarc\p$ and $\filvardenom\p$).  The relationships between these sets of variables are stated in \cite{mcewen:thesis}.
% It may be shown trivially that the various filter variables are related by 
% \begin{displaymath}
% \filvara = \filvara\p
% \spcend ,
% \end{displaymath}
% \begin{displaymath}
% \filvarb = (2/\pnorm-1) \filvara\p + \filvarb\p
% \spcend ,
% \end{displaymath}
% \begin{displaymath}
% \filvarc = (2/\pnorm-1)^2 \filvara\p + 2(2/\pnorm-1)\zreal(\filvarb\p) + \filvarc\p
% \spcend ,
% \end{displaymath}
% and 
% \begin{displaymath}
% \filvardenom = \filvardenom\p
% \spcend .
% \end{displaymath}
For now we simply note that all filter variables are identical for $\pnorm=2$, however the case $\pnorm=1$ is adopted by \cite{schaefer:2004}.
Simplifying the formula for the \mf\ given by \eqn{\ref{eqn:mf}} to the azimuthally symmetric setting, one finds that the resulting expression is identical to that derived by \cite{schaefer:2004}.
% obtains the following expression for the \mf\ of a symmetric template:
% \begin{displaymath}
% \shc{(\fil_\scalepnorm)}{\el}{0}=
% \frac{\shc{\tmpl}{\el}{0}}
% {
% \filvara\p \:
% \noisecl_\el
% }
% \spcend ,
% \end{displaymath}
% which is identical to the form derived by \cite{schaefer:2004}.
Simplifying the formula for the \saf\ given by \eqn{\ref{eqn:saf}} in a similar manner, one obtains the following expression for the \saf\ of a symmetric template:
\begin{displaymath}
\shc{(\fil_\scalepnorm)}{\el}{0} = 
(\filvardenom\p \noisecl_\el)^{-1} 
\shc{\tmpl}{\el}{0} \:
\Bigl [
\fconstaz \filvarb\p{}^\cconj + \filvarc\p 
   - 
( \fconstaz \filvara\p + \filvarb\p )
\displaystyle
\frac{\dx {\rm ln} \shc{\tmpl}{\el}{0}}{\dx {\rm ln} \el}
\Bigr ]
% \label{eqn:saf_schafer}
\spcend .
\end{displaymath}
It is apparent that the \fconstaz\ coefficient of the first $\filvarb\p$ and of $\filvara\p$ are in general dependent on the choice of the \pnormtext\ preserved by the dilation.  For $\pnorm=1$, the case considered by \cite{schaefer:2004}, one finds $\fconstazo=1$, not two as given in \cite{schaefer:2004}.  When we repeat the derivation of the \saf\ for the azimuthally symmetric case given by \cite{schaefer:2004} explicitly, we also get a unity term rather than a factor of two for these coefficients, and hence correct an algebraic error in \cite{schaefer:2004}.
% , and the resulting expression for the \saf\ of an azimuthally symmetric template
% % one finds
% % \begin{equation*}
% % \shc{(\fil_\scalepnorm)}{\el}{0} =
% % (\filvardenom\p \noisecl_\el)^{-1} 
% % \shc{\tmpl}{\el}{0} \:
% % \Bigl\{ 
% % \filvarb\p{}^\cconj + \filvarc\p 
% %   -
% % \left( \filvara\p + \filvarb\p \right)
% % \displaystyle
% % \frac{\dx {\rm ln} \shc{\tmpl}{\el}{0}}{\dx {\rm ln} \el}
% % \Bigr\}
% % \spcend ,
% % \end{equation*}
% % which 
% matches the form we derive when following the framework outlined by \cite{schaefer:2004} (although % this differs to the expression given by \cite{schaefer:2004} due to a small error in their algebra). 
The forms derived herein for the \mf\ and \saf\ on the sphere of a directional template therefore reduce to the correct forms derived directly for an azimuthally symmetric template.
Moreover, in the flat, continuous limit, the azimuthally symmetric optimal filters derived on the sphere reduce to the forms derived previously on the plane
% (we consider the azimuthally symmetric filters only, since the directional \saf\ has not yet been derived on the plane).
(to make a comparison see, respectively, \eg\ eqn.~(21) of \cite{haehnelt:1995} and eqn.~(10) of \cite{sanz:2001}).
% ; see \cite{mcewen:thesis} for a more detailed discussion of this approximation).
% However, note that for the \saf\ the coefficients of some filter variables do not in general depend on the dimension of the space, but rather on the choice of norm that the dilation is defined to preserve. 

%=============================================================================
\subsection{Comparison}
\label{sec:filters_comparison}

The relative merits of the \mf\ and \saf\ are compared in this subsection.
Recently, the advantages of the \saf\ (on the plane) have been questioned \cite{vio:2004}, although these concerns have been refuted by the original proponents of the \saf\ \cite{sanz:2001,herranz:2002}.  We hope to clarify this debate by suggesting that in the ideal case
the \saf\ may not provide a theoretical advantage,  however in practice the \saf\ may indeed prove advantageous.
%it may be true that the \saf\ provides no advantage, however in practice the \saf\ may indeed prove advantageous.

The \saf\ filter imposes an extremum in scale in the filtered field and thus must satisfy an additional constraint to the \mf.  Consequently, the gain of the \saf\ must be lower than that of the \mf.  One may show this analytically by comparing the ratio of the detection levels derived in the preceding sections for each optimal filter:
\begin{equation}
\label{eqn:gain_ratio}
\frac{\detect_{\rm \saf}}{\detect_{\rm \mf}}
=
\sqrt{1-\frac{| \filvarb |^2}{\filvara \filvarc}}
\spcend .
\end{equation}
Filter variables \filvara\ and \filvarc\ are real and are strictly positive always, consequently \eqn{\ref{eqn:gain_ratio}} is less than one always.

As the gain of the \saf\ is lower than the gain of the \mf\ one would hope that the \saf\ provides some other advantage.  This is indeed the case when the scale of the source template is unknown.  When the source size is unknown the observed field may be filtered at a range of candidate scales.   It is then possible, at the source
position, to trace the value of the filtered field over scales.  The \saf\ imposes a peak (with respect to scale) in the filtered field that may then be related, either analytically or numerically, to the unknown size of the source template.
% For the \saf\ the position of the peak in the filtered field may then be related, either analytically or numerically, to the unknown size of the source template.
% 
% For the \saf\ it is then possible to relate, either analytically or numerically, the position of the peak in the filtered field to the unknown scale of the source template.  Peaks in scale in the filtered field may then be used to estimate the unknown size of the template profile.  
The drop in gain provided by the \saf\ is therefore offset by the ability to determine the unknown size of the source.

It has been argued that it is also possible to estimate an unknown template size using the \mf\ \cite{vio:2004}, in which case the \saf\ would not provide any advantage.  For the \mf, it may be possible to relate the filtered field curve (with respect to scale) to the unknown scale of the template.  By fitting the entire curve one could therefore directly estimate the source scale, although this curve is unlikely to have a local peak.
% In ideal circumstances the \saf\ would not provide any advantage, however this is not necessarily the case in practice.  
In practice it is much easier to determine a peak in the filtered field (the \saf\ approach to estimating source size), than try to fit a curve to the field (the \mf\ approach to estimating source size).  The amplitude of the source is unknown also, hence it is only the curvature of the curve that one may fit and curvature changes much more rapidly about an extremum.  
It is therefore clear that the \saf\ does indeed provide some practical advantages to the \mf\ when the scale of the source profile is unknown.

%=============================================================================
\section{Simulations}
\label{sec:simulations}
%=============================================================================

We apply the optimal filter theory derived in the preceding section to the detection of compact objects from simulated data.  A very simple detection strategy is adopted.  The development of more sophisticated detection strategies, a rigorous quantification of the performance of the resulting object detection and the application to real data is the subject of future work.  The motivation here is to demonstrate the theory with a very simple example only.  

%=============================================================================
\subsection{Optimal filters}
\label{sec:sims_filters}

\begin{figure*}
\begin{minipage}{\textwidth}
\centering
\mbox{
\subfigure[Butterfly template]{\includegraphics[width=\plotwidth]{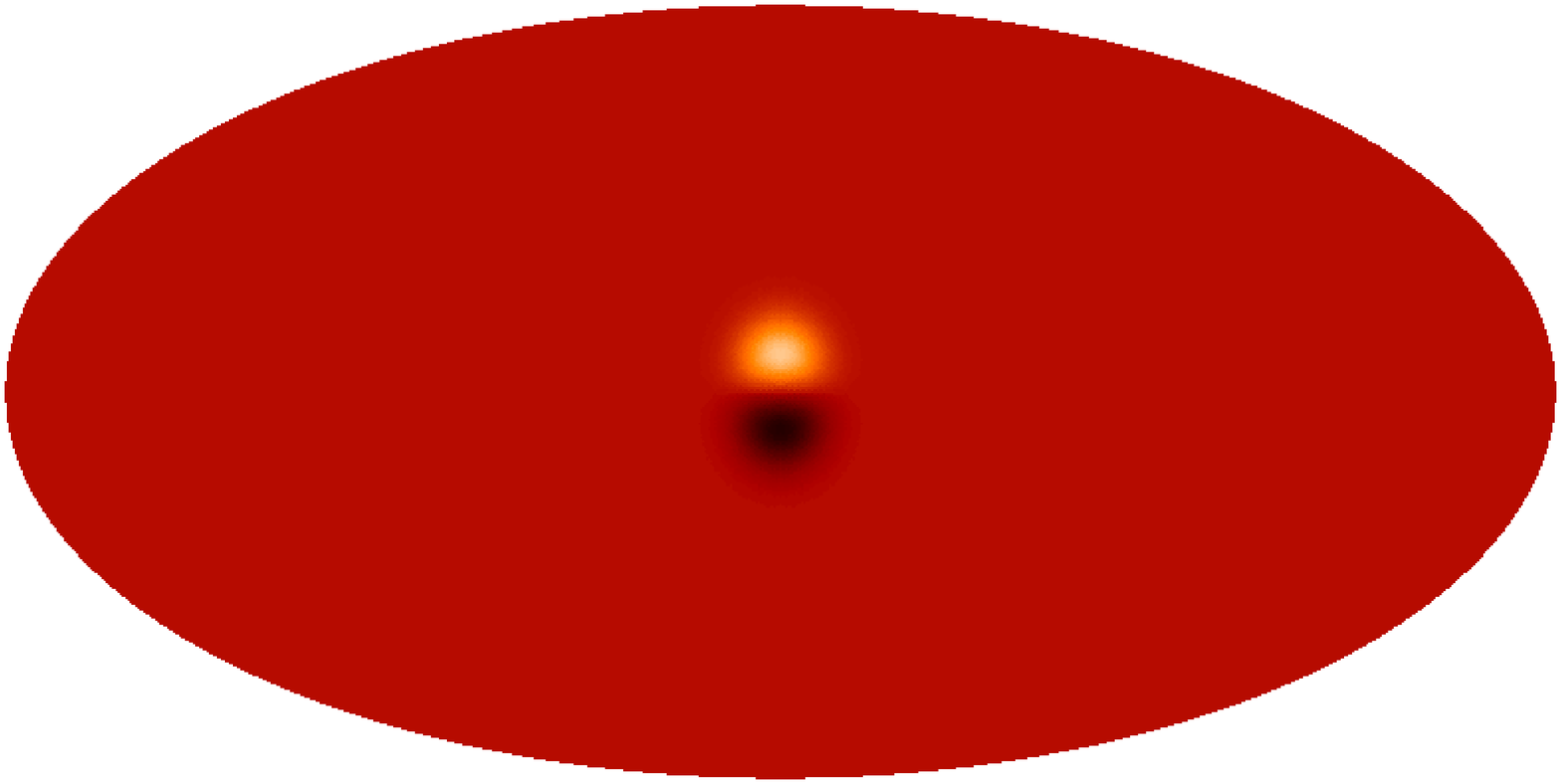}} \quad
\subfigure[\mf]{\includegraphics[width=\plotwidth]{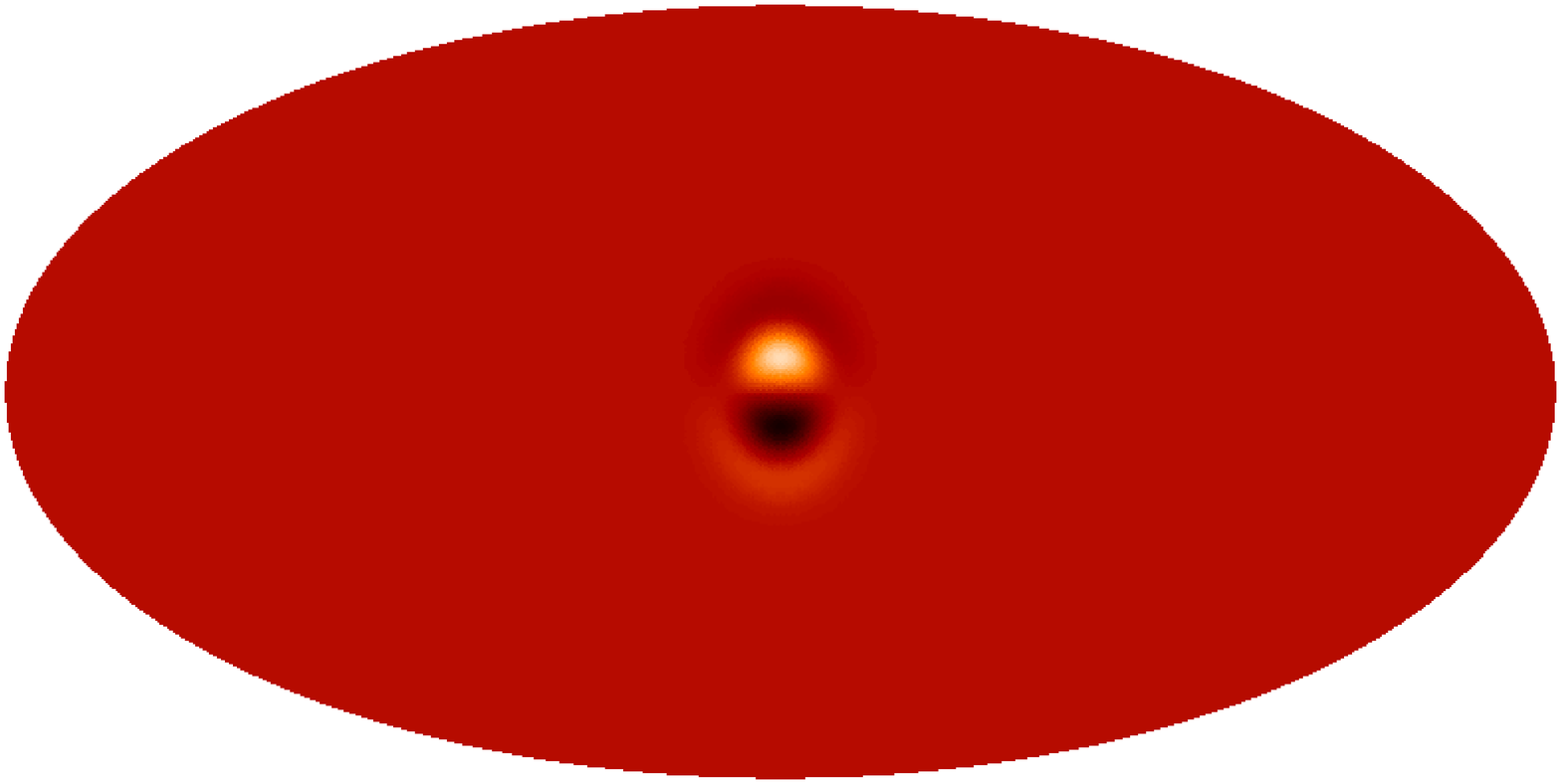}} \quad
\subfigure[\saf]{\includegraphics[width=\plotwidth]{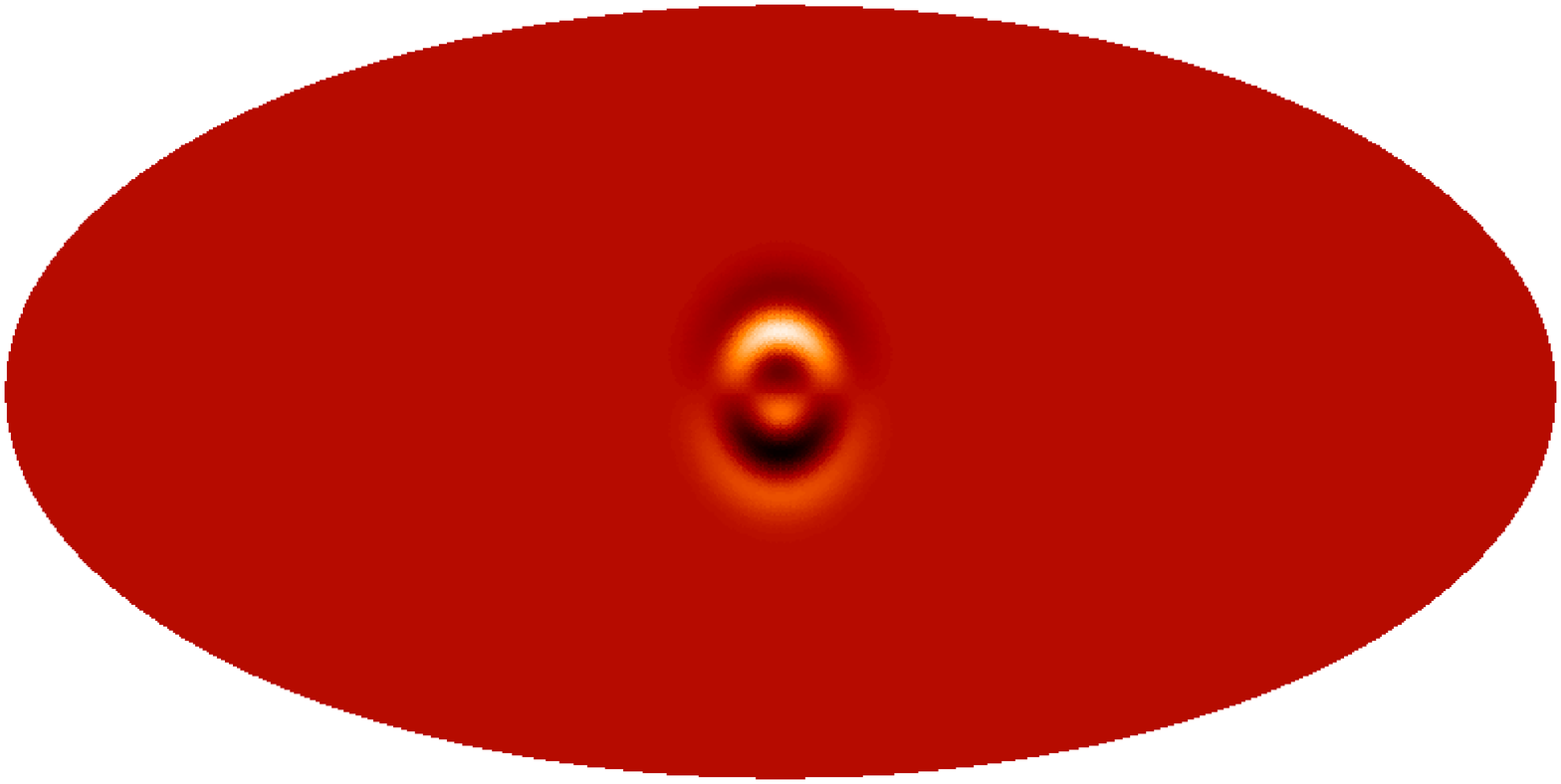}}
}
\caption{Optimal filters for the butterfly template constructed on the sphere  in the setting consistent with \wmap\ observations (see text).  (Functions/data defined on the sphere are illustrated here and subsequently using the Mollweide projection.)}

\label{fig:filters}
\end{minipage}
\end{figure*}

We construct examples of optimal filters defined on the sphere using the filter expressions derived in \sectn{\ref{sec:filters}}.  Our ultimate aim is to apply filters to detect compact objects embedded in \cmb\ data, such as the recent observations made by the NASA \wmaptext\ (\wmap) \cite{hinshaw:2006}.
Optimal filters are therefore constructed here in an analogous simulated setting. The \cmb\ power spectrum that best fits the \wmap\ data is used to model the stochastic background process, with a Gaussian beam of full-width-half-maximum (\fwhm) of $13\arcmin$ applied.  Isotropic white noise of constant variance $0.05({\rm mK})^2$ is added to reflect the noise in \wmap\ observations.
We consider the construction in this setting of the \mf\ and \saf\ for the directional butterfly template illustrated in \fig{\ref{fig:filters}~(a)} (the butterfly template is defined by the partial derivative in one direction only of a two-dimensional Gaussian on the sphere; see \cite{mcewen:2006:fcswt} for a definition).\footnote{The step of the butterfly template may be used to model the line-like discontinuity of Kaiser-Stebbins type cosmic string signatures \cite{kaiser:1984}.}  The resultant \mf\ and \saf\ are illustrated in \fig{\ref{fig:filters}~(b) and (c)} respectively.  Optimal filters are constructed here in the context of \ltwo-norm preserving dilations, \ie\ for $\pnorm=2$.  Furthermore, in practice the template profiles are assumed to be band-limited at \elmax, in which case all expressions and summations involving \el\ are computed up to \elmax\ only.  In these experiments we choose $\elmax=256$ since this is a more than adequate band-limit to ensure that all structure of the butterfly profile is considered.

A Fortran 90 package called
\stwofil\footnote{\label{footnote:download}The \stwofil, \fastcswt\
  and \comb\ packages are available for download from \url{http://www.mrao.cam.ac.uk/~jdm57/}.  These packages require the \healpix\ (\url{http://healpix.jpl.nasa.gov/}) \cite{gorski:2005} and {\tt FFTW} (\url{http://www.fftw.org/}) libraries.} (\sphere\ FILtering) has been implemented to compute the equations describing the \mf\ and \saf\ defined by \thmref{\ref{thm:mf}} and \thmref{\ref{thm:saf}} respectively.  This package makes use of our \fastcswt${}^{\mbox{\scriptsize\ref{footnote:download}}}$ package \cite{mcewen:2006:fcswt} to perform fast filtering on the sphere.
Additional numerical considerations must be taken into account when implementing the filter equations since they often involve dividing the template spherical harmonics by the power spectrum of the background process, which becomes problematic when taking the ratio of two very small numbers.  These considerations are discussed in detail in \cite{mcewen:thesis}.  

% \begin{figure}
% \centering
% \subfigure[Butterfly template]{\includegraphics[width=\plotwidth]{figures/mf_Tbfly_Bcmb_Stmpl_tmpl_pm_red}} 
% \subfigure[\mf]{\includegraphics[width=\plotwidth]{figures/mf_Tbfly_Bcmb_Stmpl_sky01_pm_red}}
% \subfigure[\saf]{\includegraphics[width=\plotwidth]{figures/saf_Tbfly_Bcmb_Stmpl_sky01_pm_red}}
% \caption{Optimal filters for the butterfly template constructed on the sphere  in the setting consistent with \wmap\ observations (see text).  (Functions/data defined on the sphere are illustrated here and subsequently using the Mollweide projection.)}
% \label{fig:filters}
% \end{figure}

%=============================================================================
\subsection{Object detection}

% We demonstrate in this subsection the detection on the sphere of compact objects embedded in a stochastic background process.  We are not concerned with quantifying accurately the performance of the detection strategy adopted but merely demonstrate the application of optimal filter theory on the sphere to simple object detection. 

To demonstrate optimal filter theory applied to object detection a simple example is considered.  Although the optimal filters we have derived may be used to detect objects of unknown size, in this simple demonstration we consider only a butterfly template of a fixed, known size.  Since the template size is known only the \mf\ on the sphere is applied.  The \mf\ is applied to simulations of the \cmb\ with artificially embedded butterfly sources, in order to recover the positions, orientations and amplitudes of the sources.  Firstly, the simulation pipeline used to construct this data is described.  The object detection procedure is then described briefly, before the results obtained from applying this procedure to simulated data are presented.

\subsubsection{Simulation pipeline}

To test the application of optimal filters on the sphere to object detection, simulated data where the ground truth is known is analysed.  We have implemented a Fortran 90 package called \comb${}^{\mbox{\scriptsize\ref{footnote:download}}}$ (COmpact eMBedded object simulations) to facilitate such simulations.
\comb\ allows the user to embed compact functions on the sphere within a stochastic background process.  The parameters of the embedded objects are uniformly sampled from some prescribed interval, whereas the background process is specified by its power spectrum.  Functionality is also included to incorporate noise and beams.  

An example of maps simulated using \comb\ is shown in the first four panels of \fig{\ref{fig:comb_sims}}.  In this example the background process is described by the best-fit \wmap\ \cmb\ power spectrum, a beam of $13\arcmin$ \fwhm\ is applied and isotropic white noise of variance 0.05(mK)${}^2$ is added (the same setting in which the optimal filters discussed in \sectn{\ref{sec:sims_filters}} were constructed).  A map of simulated butterfly sources is shown in \fig{\ref{fig:comb_sims}~(c)} and is embedded in the resultant simulated map shown in \fig{\ref{fig:comb_sims}~(d)}.  
In this situation the signal-to-noise ratio (\snr) of an embedded object is defined by the ratio of the peak amplitude of the template relative to the root-mean-squared (RMS) value of the \cmb\ background.
For the butterfly templates it can be shown that ${\rm \snr} \approx 0.4 \amp$, where \amp\ is the amplitude of the object (note that the butterfly template definition is normalised, with a peak amplitude of 0.13mK for $\amp=1.0$). 
In the example illustrated in \fig{\ref{fig:comb_sims}} all objects have a constant amplitude of $\amp=1.0$, corresponding to ${\rm \snr}=0.40$.  Moreover, in this simulation all embedded objects have the same orientation ($\eulc=0^\circ$); that is, the orientation of the objects are aligned to point to the north pole.  This is a useful test case since it reduces the complexity of the subsequent object detection to that of a symmetric template profile, while still adopting the more general optimal filter definitions.  In the following object detection we also consider simulations where the orientations of the objects vary, however object orientations are only allowed to lie on a discrete, uniformly sampled grid with $N_\eulc=5$ samples.  By restricting the allowable orientations to a grid it is necessary only to compute the filtered field for a small number of orientations.  Obviously in practice object orientations will not lie on a grid, however in this case one may simply compute the filtered field for a higher orientation resolution (or one may use steerable filters\footnote{Steerable template functions yield steerable optimal filters on the sphere since the filter equations do not mix \m\ structure.  See \cite{wiaux:2005} for a discussion of steerable filters on the sphere.} to extract the orientation corresponding to the most dominant feature, thereby effectively considering all continuous orientations; this is the topic of future work).
An orientational resolution of $N_\eulc=5$ is sufficient for the simple demonstration presented here.

\subsubsection{Detection procedure}
\label{sec:detect_proc}

For the purpose of this demonstration we perform only a naive detection strategy based on {thres\-holding} the filtered field.  Firstly the initial map is filtered using the \mf\ to enhance the contribution due to the embedded sources relative to the background.   The local maxima of the filtered field that lie above a certain threshold are used to define detected objects.  The filtered field is thresholded at a level determined by a constant ($N_\sigma$) times the standard deviation of the map.  Typically an $N_\sigma$ of 2.5 or 3.0 is used.  
The magnitude and position of each local maximum that remains in the thresholded field is used to compute the parameters of detected sources.  This object detection procedure is implemented in the \stwofil\ package.  
% In this simple demonstration objects of a fixed size are considered only, however the detection strategy may be extended easily to account for objects of varying size.

\begin{figure*}
\begin{minipage}{\textwidth}
\centering
\mbox{
\subfigure[\cmb]{\includegraphics[width=\plotwidth]{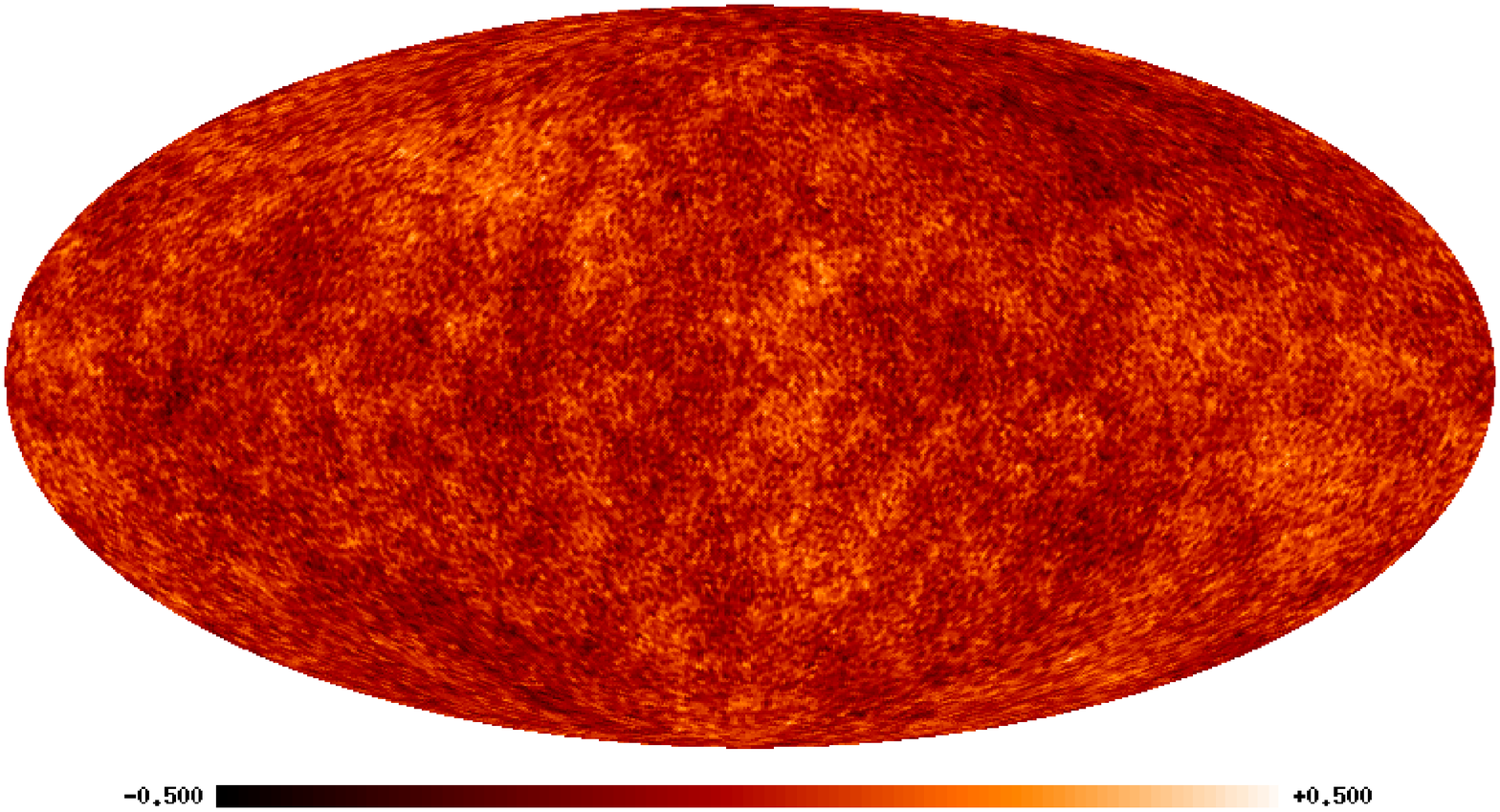}} \quad
\subfigure[White noise]{\includegraphics[width=\plotwidth]{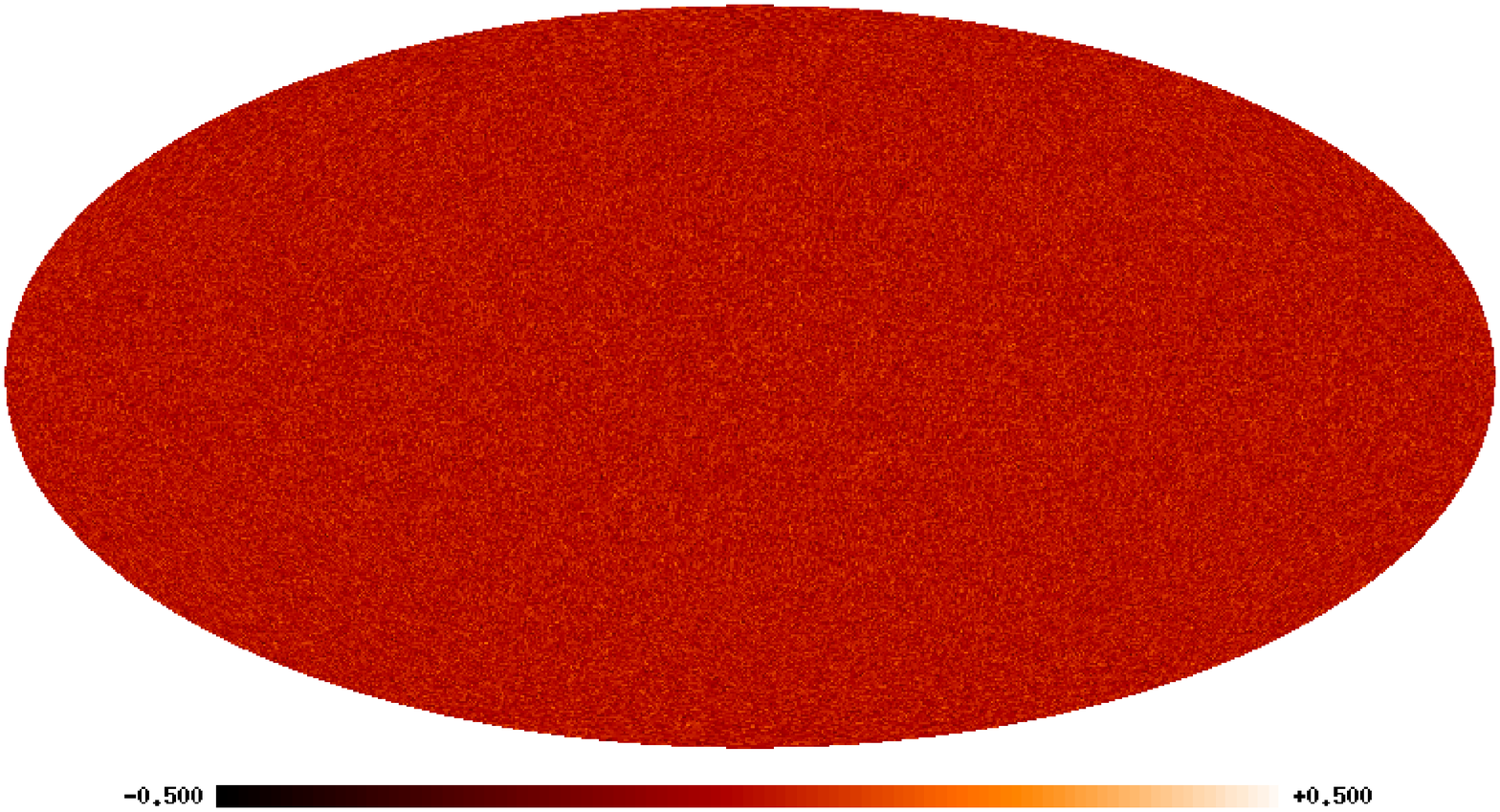}} \quad
\subfigure[Embedded objects]{\includegraphics[width=\plotwidth]{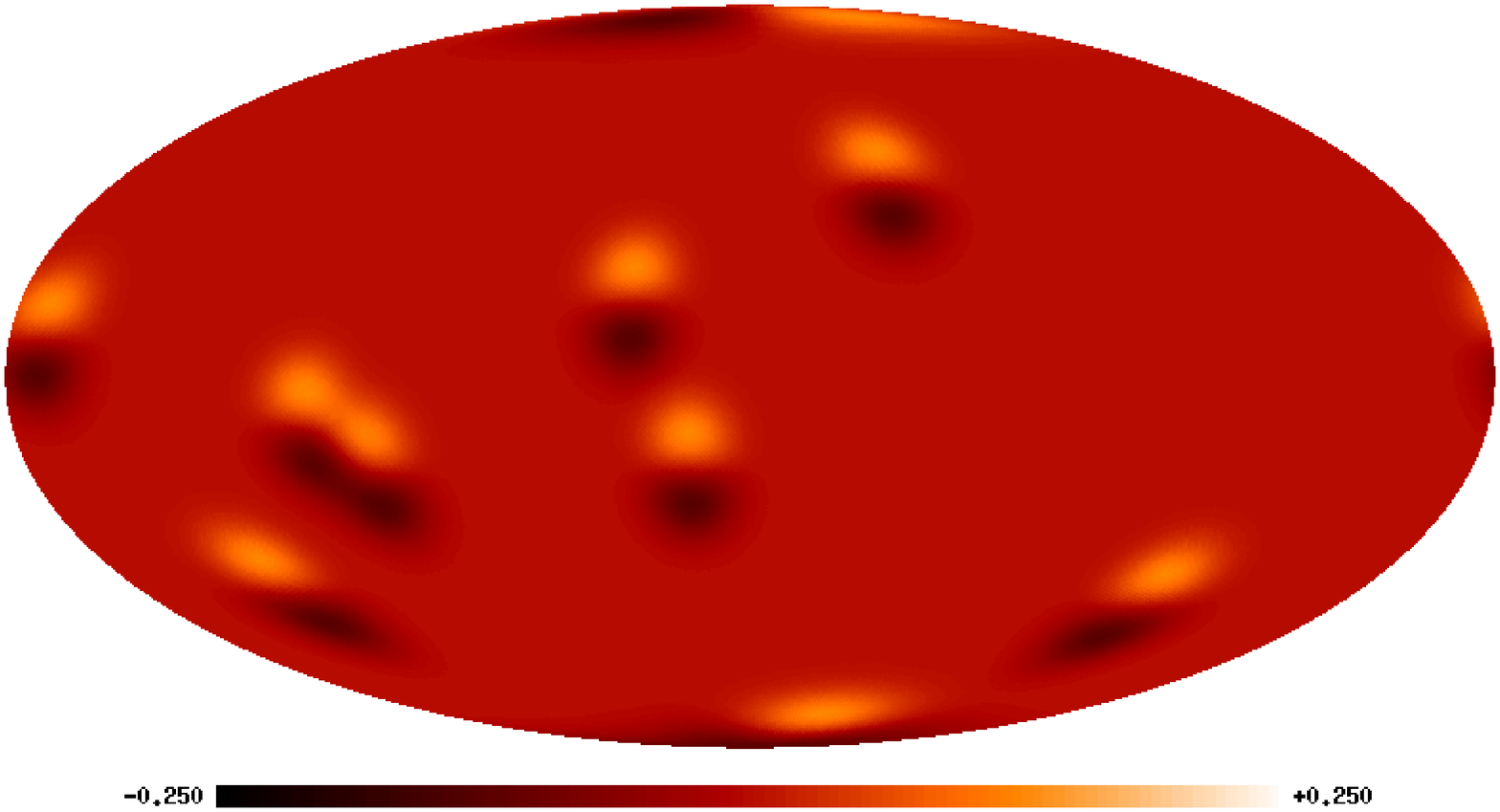}}
}
\mbox{
\subfigure[Simulated sky with objects embedded]{\includegraphics[width=\plotwidth]{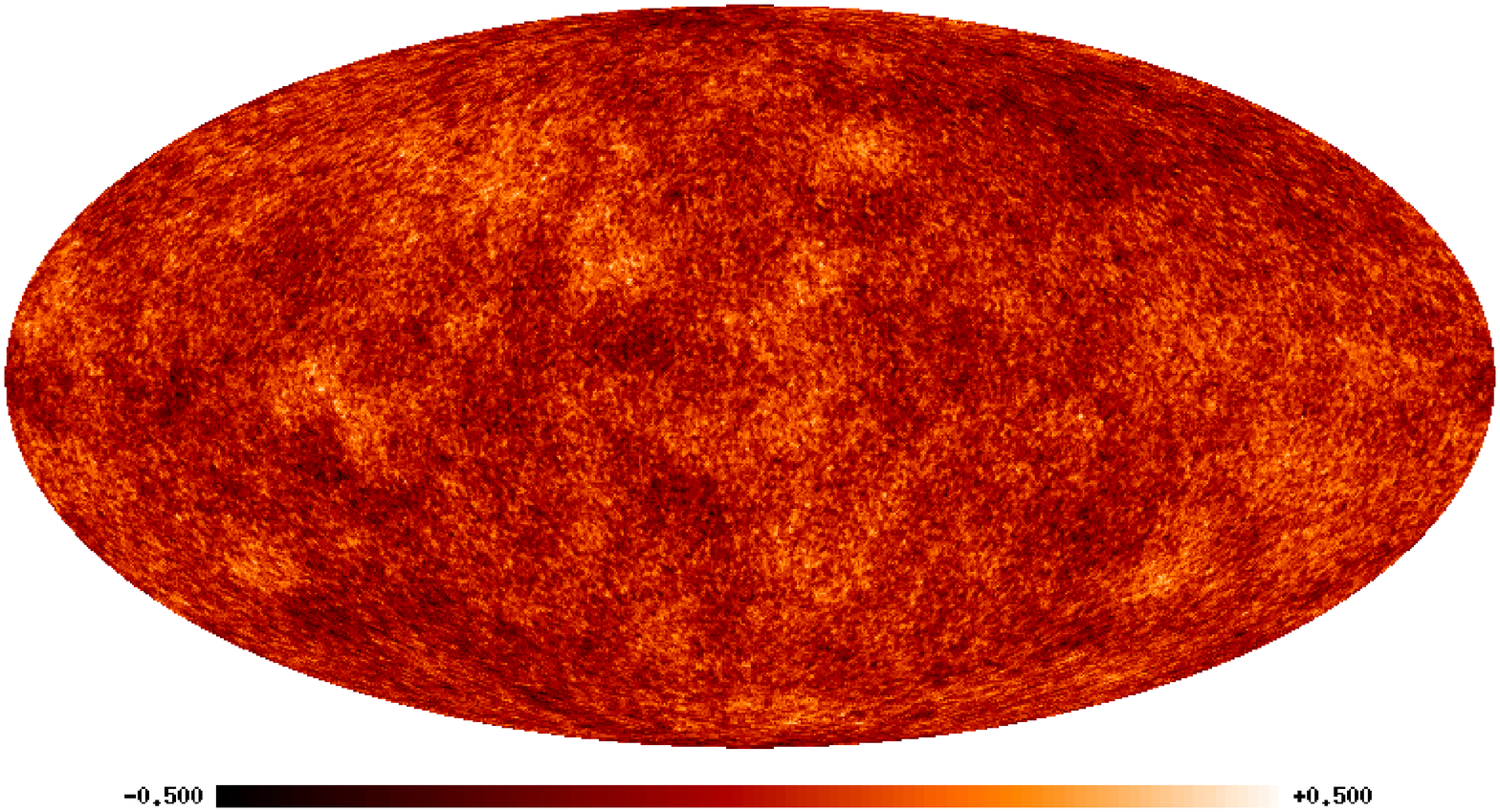}} \quad
\subfigure[Filtered field]{\includegraphics[width=\plotwidth]{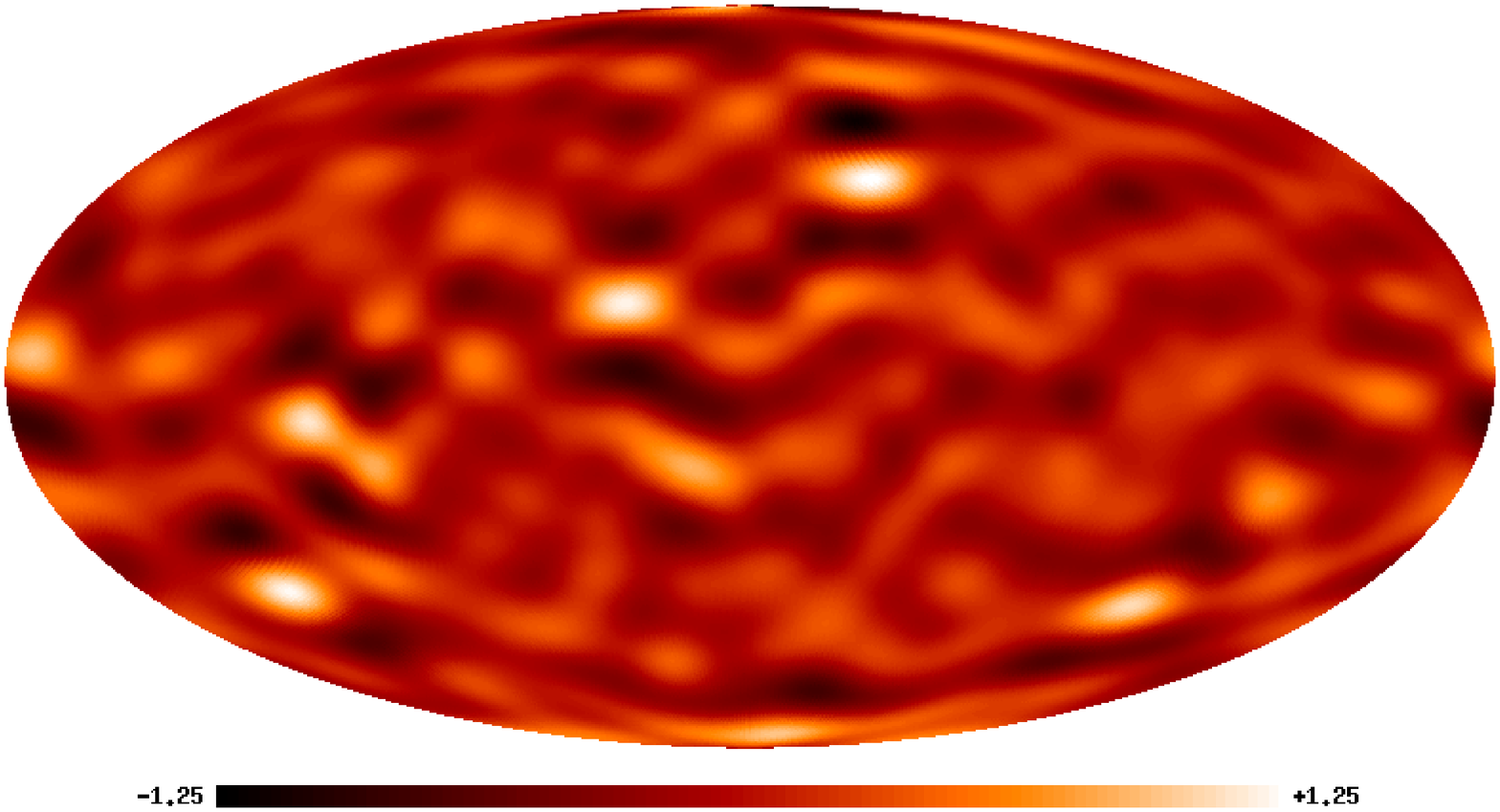}} \quad
\subfigure[Recovered objects]{\includegraphics[width=\plotwidth]{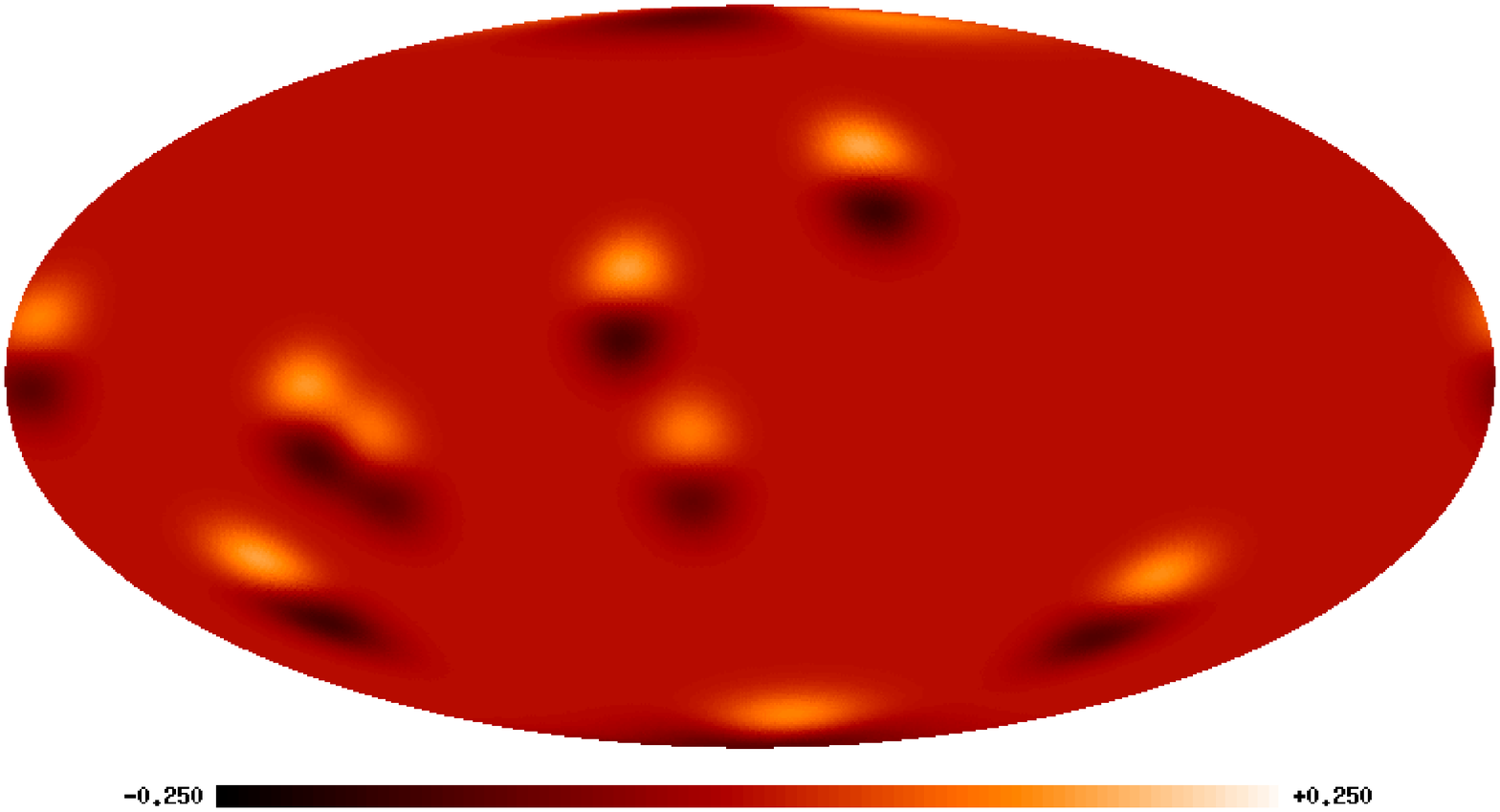}}
}
\caption{Embedded object simulation and recovered objects.  
The \cmb\ background shown in panel~(a) is simulated in accordance with the \cmb\ power spectrum consistent with current observations, noise of variance 0.05(mK)${}^2$ (panel~(b)) is added and a beam of $13\arcmin$ \fwhm\ is applied.  The simulated butterfly templates (${\rm \snr}=0.40$) shown in panel~(c) are embedded in the the resultant simulated map shown in panel~(d).  The \mf\ filtered field of the simulated sky is shown in panel~(e).  The local maxima in the filtered field are detected by thresholding at $N_{\sigma}=3.0$ to recover the compact objects depicted in panel~(f).  
In this example all embedded objects are detected and no false detections are made.
(All maps shown here and subsequently are displayed in units of mK.)
}
\label{fig:comb_sims}
\end{minipage}
\end{figure*}

This detection algorithm is applied to the simulated data illustrated in \fig{\ref{fig:comb_sims}}.  The filtered field is displayed in \fig{\ref{fig:comb_sims}~(e)} and the objects recovered from the local maxima of the filtered field are shown in \fig{\ref{fig:comb_sims}~(f)}. 
% The parameters of the embedded and recovered objects are displayed in \tbl{\ref{tbl:detect1}}.  Errors on the amplitude estimates are calculated from the variance of the filtered field at the source position.  
Notice that for this simulation at ${\rm \snr}=0.40$ all embedded objects are recovered accurately and no false detections are made.  We now consider more difficult object detection problems.

\subsubsection{Results}

The example illustrated in \fig{\ref{fig:comb_sims}} demonstrates object detection on the sphere for a relatively easy case.  By considering more difficult detection situations we examine the limits of the simple detection algorithm described above.  Firstly, only a single fixed orientation is considered, before the orientation of embedded objects is allowed to vary.

For objects of known orientation we examine the effect of reducing the \snr\ of the embedded sources on the number of successful and false detections.  We also experiment with thresholding levels \mbox{$N_\sigma=3.0$} and $N_\sigma=2.5$.  The results of these tests are given in \tbl{\ref{tbl:detections}}.
To ensure minimal false detections are made one should threshold at $N_\sigma=3.0$, however to improve the completeness of positive detections one may use $N_\sigma=2.5$, at the expense of a small number of false detections.
% 
% We conclude the tests for sources of a known orientation by demonstrating the extraction of objects at a range of different amplitudes and state the parameter estimates obtained.  One sigma errors on the amplitude estimates are calculated from the variance of the filtered field at the source position (as discussed in \sectn{\ref{sec:field_stats}}).  The actual and recovered object parameters are specified in \tbl{\ref{tbl:detect2}} and may be observed in \fig{\ref{fig:sim3}}.  

\begin{table}
\caption{Object detection performance for $\eulc=0\degrees$}
\label{tbl:detections}
\centering
\scriptsize
\begin{tabular}{ccccc} \toprule
  \snr & $N_\sigma$ & \multicolumn{2}{c}{Detections} & Number of sources\\
       &            & Correct            & False            \\ \midrule
  0.40  & 3.0        & 10                 & 0          & 10  \\
  0.40  & 2.5        & 10                 & 1          & 10  \\
%   0.30  & 3.0        &  7                 & 0          & 10  \\
%   0.30  & 2.5        & 10                 & 1          & 10  \\
  0.25  & 3.0        &  6                 & 0          & 10  \\
  0.25  & 2.5        &  8                 & 2          & 10  \\
  0.20  & 3.0        &  5                 & 0          & 10  \\
  0.20  & 2.5        &  7                 & 6          & 10  \\ \bottomrule
\end{tabular}
\end{table}

\begin{figure*}
\begin{minipage}{\textwidth}
\centering
\mbox{
\subfigure[Simulated sky with objects embedded]{\includegraphics[width=\plotwidth]{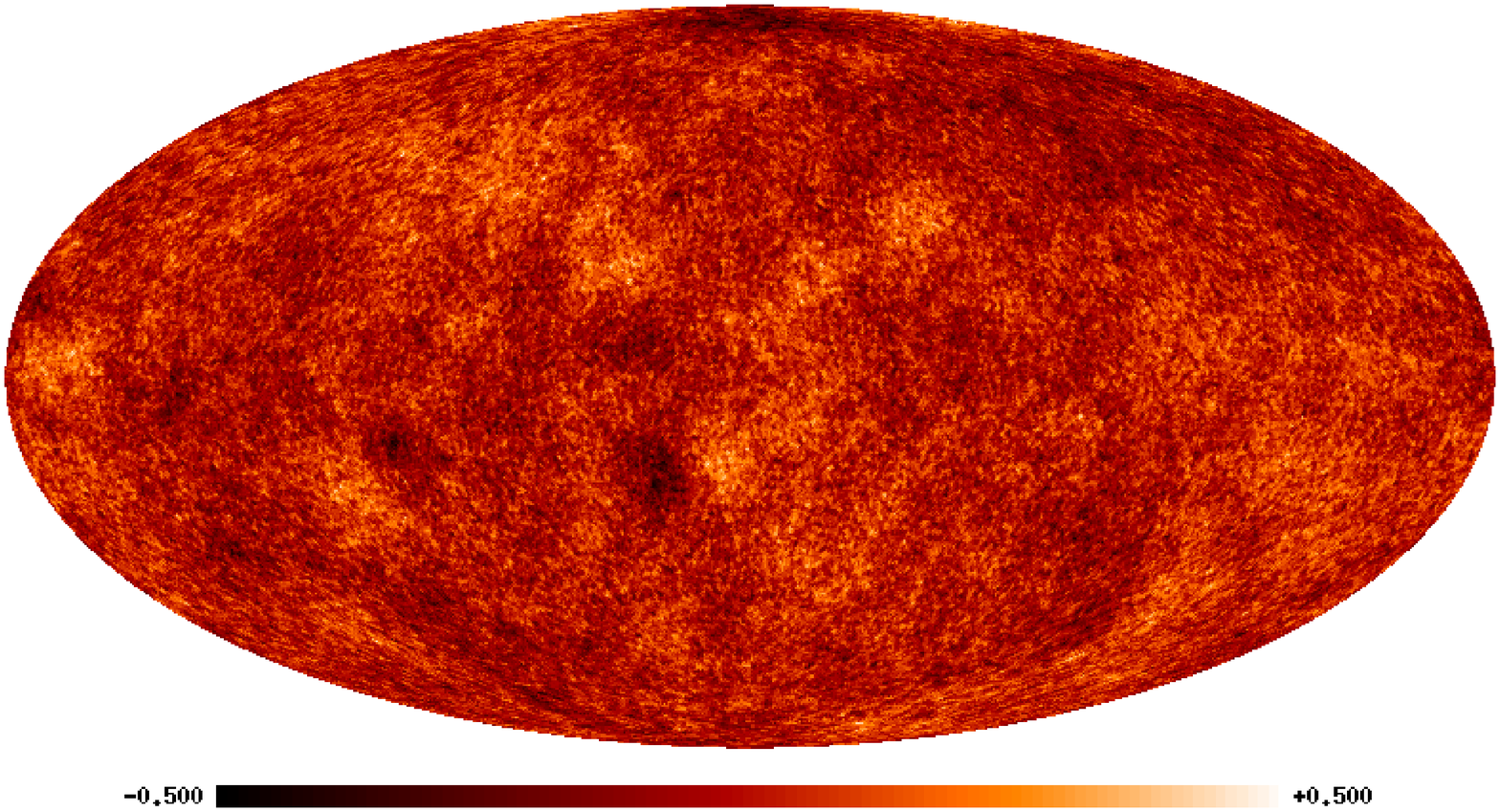}} \quad
\subfigure[Actual objects]{\includegraphics[width=\plotwidth]{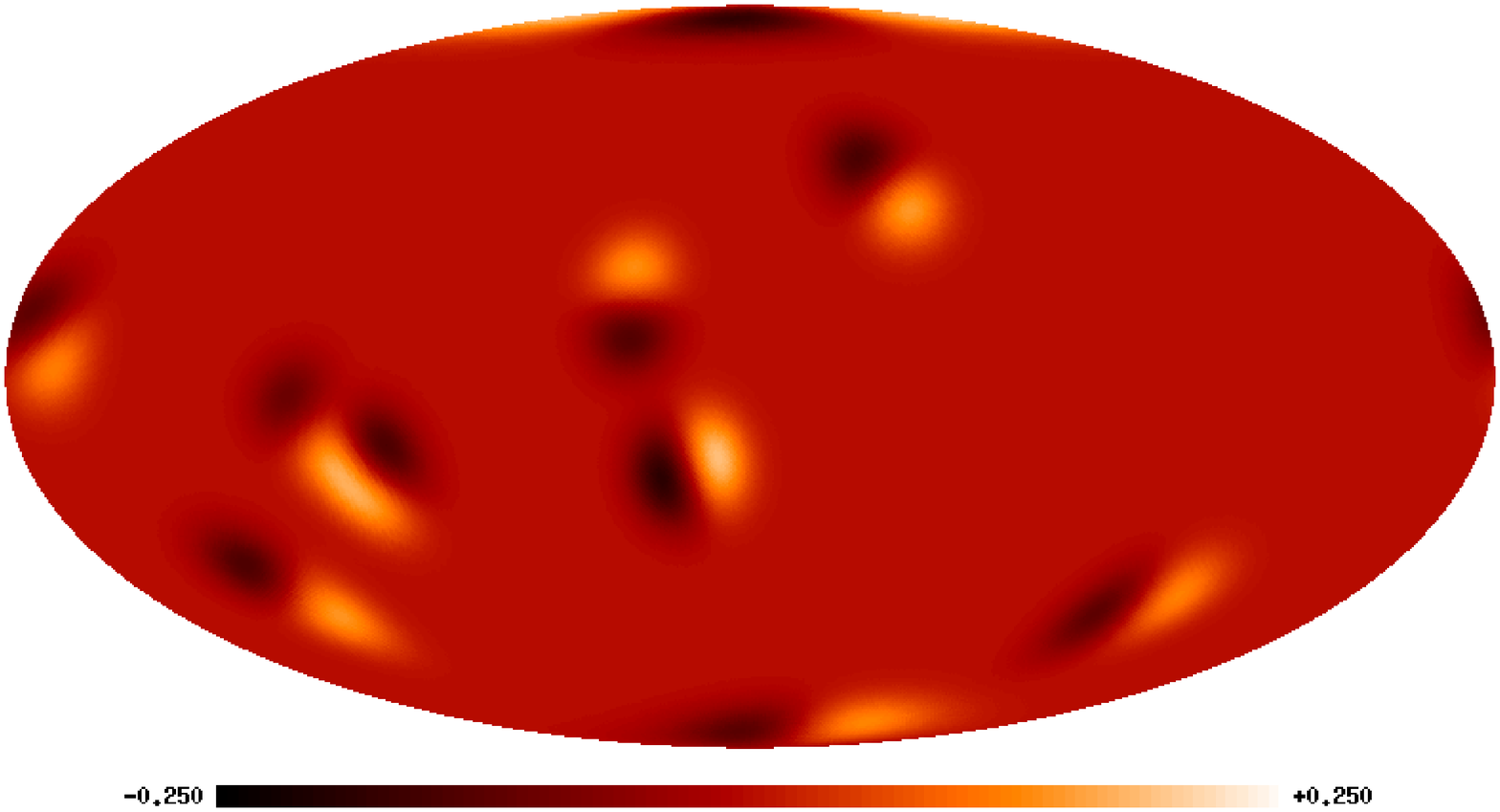}} \quad
\subfigure[Recovered objects]{\includegraphics[width=\plotwidth]{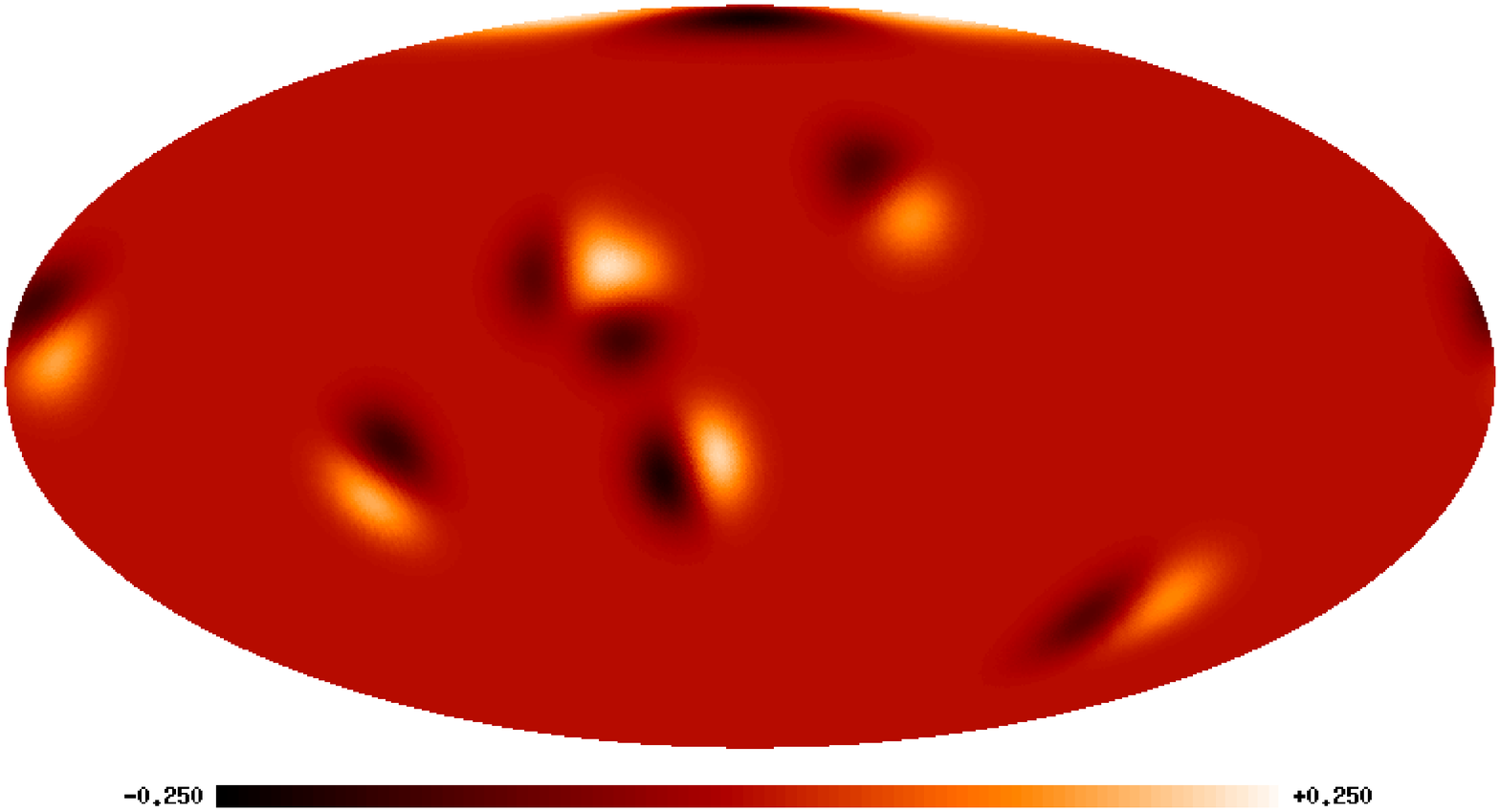}}
}
\caption[Embedded object simulation (range of \snr s; $N_{\rm \gamma}=5$; $N_{\sigma}=2.5$)]{Embedded object simulation and recovered objects for a range of \snr s, $N_{\rm \gamma}=5$ and $N_{\sigma}=2.5$.  See \tbl{\ref{tbl:detect4}} for the actual and recovered parameters of the embedded sources.}
\label{fig:sim6}
\end{minipage}
\end{figure*}

We now consider objects with an unknown orientation laying on a uniform grid of resolution $N_\eulc=5$.  Allowing orientations to vary introduces an extra degree-of-freedom and thus makes the object detection problem more difficult.  At $\snr=0.40$, for thresholding levels $N_\sigma=3.0$ and $N_\sigma=2.5$, of ten embedded objects we recover six and eight objects correctly and make zero and one false detection respectively.  At $\snr=0.25$ we recover five objects correctly and make one false detection for $N_\sigma=3.0$ ($N_\sigma=2.5$ is not appropriate for this low \snr).
Finally, we demonstrate in this setting the recovery of objects at a range of different amplitudes. 
The actual and recovered object parameters are described in \tbl{\ref{tbl:detect4}} and may be observed in \fig{\ref{fig:sim6}}.
%%%%
One sigma errors on the amplitude estimates are calculated from the variance of the filtered field at the source position (as discussed in \sectn{\ref{sec:field_stats}}). 
%%%%

It is important to note that the detection strategies demonstrated here are extremely naive and are presented merely to demonstrate the new filter framework derived on the sphere.  A more rigorous approach is to perform more sophisticated detection classification schemes, such as the Neyman-Pearson test.  Moreover, for certain classes of template functions which are steerable, one may use steerable filters to extract a single orientation over the domain of all continuous orientations.  This is expected to improve considerably the performance of object detection in cases of varying source orientation, to the extent that one would expect the performance to match that of cases where the source orientation is known.  These extensions are the focus of future work.

% \begin{figure}
% \centering
% \subfigure[Simulated sky with objects embedded]{\includegraphics[width=\plotwidth]{figures/ngamma5/bfly_csky_sc_red}}
% \subfigure[Actual objects]{\includegraphics[width=\plotwidth]{figures/ngamma5/bfly_obj_sc_red}}
% \subfigure[Recovered objects]{\includegraphics[width=\plotwidth]{figures/ngamma5/bfly_obj_loc_sc_red}}
% \caption[Embedded object simulation (range of \snr s; $N_{\rm \gamma}=5$; $N_{\sigma}=2.5$)]{Embedded object simulation and recovered objects for $N_{\rm \gamma}=5$ and $N_{\sigma}=2.5$.  See \tbl{\ref{tbl:detect4}} for the actual and recovered parameters of the embedded sources.}
% \label{fig:sim6}
% \end{figure}

\begin{table}
\centering
\caption{Actual and recovered object parameters}
\label{tbl:detect4}
\scriptsize
\begin{tabular}{cll} \\ \toprule
Source & \multicolumn{2}{c}{Object parameters} \\ 
& \multicolumn{1}{c}{Amplitude} & \multicolumn{1}{c}{Euler angles}\\\midrule

 1 & $\amp   =  1.48$          & $(\euls)   =( 15.0^\circ,108.3^\circ, 72.0^\circ)$ \\&
    $\amphat=  1.66\pm 0.20$   &$(\eulshat)=( 14.7^\circ,107.9^\circ, 72.0^\circ)$ \\

 2 & $\amp   =  1.06$           &$(\euls)   =( 29.2^\circ, 75.4^\circ,  0.0^\circ)$\\&
    $\amphat=  1.26\pm 0.20$   &$(\eulshat)=( 30.9^\circ, 75.6^\circ,  0.0^\circ)$ \\

 3 & $\amp   =  1.43$           &$(\euls)   =( 78.4^\circ,  0.7^\circ,288.0^\circ)$\\&
    \multicolumn{1}{l}{Not recovered} \\

 4 & $\amp   =  1.19$           &$(\euls)   =( 94.4^\circ,109.1^\circ,216.0^\circ)$\\&
     $\amphat=  1.34\pm 0.20$   &$(\eulshat)=( 92.6^\circ,109.3^\circ,216.0^\circ)$ \\

 5 & $\amp   =  0.81$           &$(\euls)   =(107.8^\circ, 99.6^\circ,144.0^\circ)$\\&
     \multicolumn{1}{l}{Not recovered} \\

 6 & $\amp   =  1.13$           &$(\euls)   =(136.0^\circ,133.9^\circ,144.0^\circ)$\\&
     \multicolumn{1}{l}{Not recovered} \\

 7 & $\amp   =  0.93$           &$(\euls)   =(172.3^\circ, 82.7^\circ,144.0^\circ)$\\&
    $\amphat=  1.25\pm 0.20$   &$(\eulshat)=(171.9^\circ, 81.2^\circ,144.0^\circ)$ \\

 8 & $\amp   =  0.95$           &$(\euls)   =(241.2^\circ,137.2^\circ, 72.0^\circ)$\\&
     $\amphat=  1.02\pm 0.20$   &$(\eulshat)=(242.8^\circ,138.0^\circ, 72.0^\circ)$ \\

 9 & $\amp   =  1.00$           &$(\euls)   =(317.7^\circ,172.3^\circ, 72.0^\circ)$\\&
    $\amphat=  1.76\pm 0.20$   &$(\eulshat)=(287.0^\circ,  180.0^\circ, 72.0^\circ)$ \\

10 & $\amp   =  1.16$           &$(\euls)   =(321.7^\circ, 50.7^\circ,144.0^\circ)$\\&
     $\amphat=  1.08\pm 0.20$   &$(\eulshat)=(321.4^\circ, 52.5^\circ,144.0^\circ)$ \\

11 & \multicolumn{1}{l}{Not present}\\&
     $\amphat=  0.96\pm 0.20$   &$(\eulshat)=( 47.0^\circ, 67.9^\circ, 72.0^\circ)$ \\

\bottomrule

\end{tabular}
\end{table}

\section{Concluding Remarks}
\label{sec:conclusions}
%=============================================================================

We have extended the concept of optimal filtering to a spherical manifold.  Expressions for the spherical \mf\ and \saf\ have been derived for general non-azimuthally symmetric template objects.  For the special case of an azimuthally symmetric template we have shown that the general results derived herein reduce to the forms derived directly in this setting.  Moreover, we have also shown that in the flat approximation the optimal filters derived on the sphere reduce to the corresponding optimal filter definitions defined on the plane.  

The focus of this paper is to derive optimal filter theory on the sphere, nevertheless we have also demonstrated the application of optimal filters to simple object detection.  We have generated simulations on the sphere of objects embedded in a stochastic background process.  Using this simulated data we have tested a simple object detection algorithm based on thresholding the optimal filtered field.  This simple object detection strategy has been demonstrated to perform well, even at low \snr.  

In the future we intend to develop more sophisticated detection classification schemes using the optimal filtered field.  Moreover, for steerable template profiles the use of steerable optimal filters is expected to improve considerably the performance of object detection in cases of varying source orientation.
We intend to then apply the resulting object detection algorithms to real \cmb\ data to search for cosmic string signatures, a theoretically postulated but as yet unobserved phenomenon \cite{kaiser:1984}.

% such as the line-like discontinuity of Kaiser-Stebbins type cosmic string effects \cite{kaiser:1984} that may be modelled by the butterfly template.
% 
% In addition, we intend to quantify rigorously the performance of the resulting object detection algorithms and apply these to real \cmb\ data to search for cosmic string signatures
% 
% \footnote{The step of the butterfly template may be used to model the line-like discontinuity of Kaiser-Stebbins type cosmic string effects \cite{kaiser:1984}.}, a theoretically postulated but as yet unobserved phenomenon \cite{davis:2005}.

%=============================================================================

\appendices
 
\section{Proof of Dilation Definition (Definition 1)}
\label{sec:appn_dilation}

We prove that the definition of an \pnormtext\ preserving
spherical dilation does indeed preserve the \pnormtext. 
The result for the case $\pnorm=2$ is given by \cite{antoine:1999} and an explicit proof is given by \cite{wiaux:2005}.
For all practical purposes only the cases 
$\pnorm=\{1,2,\infty\}$ are considered, nevertheless we prove the result for all
positive integer \pnorm, taking a different approach to that of \cite{wiaux:2005}.  First consider the case of positive
 integer $\pnorm < \infty$. % (the case $\pnorm = \infty$ is considered separately).
One requires $\|\dilp(\scale) f \|_\pnorm = \|f\|_\pnorm$, or 
equivalently $I_1=I_2$, where
% \begin{displaymath}
$
I_1 = \int_\sphere \left|f(\sas)\right|^\pnorm \dmu{\sas}
$
% \end{displaymath}
and 
% \begin{displaymath}
\mbox{$
I_2 = \int_\sphere | \left[\cocycle(\scale,\saa)\right]^{1/\pnorm} 
f(\saa_{1/\scale},\sab) |^\pnorm \dmu{\sas}
$}.
% \spcend .
% \end{displaymath}
By making a change of variables $I_2$ may be rewritten as
\begin{displaymath}
I_2 = \int_\sphere
|f(\saa,\sab)|^\pnorm
|\cocycle(\scale,\saa_\scale)|
\sin\saa_\scale 
\: \scale \:
\frac{\cos^2(\saa_\scale/2)}{\cos^2(\saa/2)}
%\frac{\cos^2 \frac{\saa_\scale}{2}}{\cos^2 \frac{\saa}{2}}
 \dx \saa \dx \sab
\spcend .
% \begin{split}
% I_2 &= \int_\sphere
% |f(\saa,\sab)|^\pnorm
% |\cocycle(\scale,\saa_\scale)|
% \sin\saa_\scale \dx \saa_\scale \dx \sab \\
% &= 
% \int_\sphere
% |f(\saa,\sab)|^\pnorm
% |\cocycle(\scale,\saa_\scale)|
% \sin\saa_\scale 
% \: \scale \:
% \frac{\cos^2 \frac{\saa_\scale}{2}}{\cos^2 \frac{\saa}{2}}
%  \dx \saa \dx \sab
% \spcend .
% \end{split}
\end{displaymath}
Thus, for $I_1=I_2$ it is apparent, after a little algebra, that 
the cocycle required to preserve the \pnormtext\ 
(for positive integer $\pnorm<\infty$) is of the 
form given by \eqn{\ref{eqn:cocycle}}.
% \begin{displaymath}
% % \begin{split}
% |\cocycle(\scale,\saa_\scale)| 
% % &=
% % \frac{\sin\saa \: \cos^2 \frac{\saa}{2} }
% % {\scale \: \sin\saa_\scale \: \cos^2 \frac{\saa_\scale}{2}} \\
% % &
% =
% \frac{4 \scale^2}
% {\left[
% (\scale^2 - 1) \cos\saa_\scale
% +(\scale^2 + 1)
% \right]^2}
% \spcend ,
% % \end{split}
% \end{displaymath}
% hence the cocycle required to preserve the \pnormtext\ 
% (for positive integer $\pnorm<\infty$) is of the 
% form given by \eqn{\ref{eqn:cocycle}}.
Now consider the \mbox{${\rm L}^\infty$-norm} case, corresponding to the case where no cocycle term is applied.
When no cocycle is applied the function is dilated only and is not scaled, hence the maximum absolute value of the function over its domain remains unaltered.  Consequently, the infinity norm is also preserved.
% One requires $\|\dil(\scalepnorm) f \|_\infty = \|f\|_\infty$.  Now
% \ifcol
%   \begin{align*}
%   \| & \dil(\scalepnorm) f \|_\infty \\ 
%   &=
%   {\rm max} \left\{
%   \left|
%   \left[\cocycle(\scale,\saa)\right]^{1/\infty}
%   f(\saa_{1/\scale},\sab)
%   \right|
%   : \theta \in [0,\pi], \sab \in [0,2\pi)
%   \right\} \\
% %   &=
% %   {\rm max} \left\{
% %   \left|
% %   f(\saa_{1/\scale},\sab)
% %   \right|
% %   : \theta \in [0,\pi], \sab \in [0,2\pi)
% %   \right\} \\
%   &= \| f \|_\infty
%   \spcend .
%   \end{align*}
% \else
%   \begin{displaymath}
%   \begin{split}
%   \|\dil(\scalepnorm) f \|_\infty &=
%   {\rm max} \left\{
%   \left|
%   \left[\cocycle(\scale,\saa)\right]^{1/\infty}
%   f(\saa_{1/\scale},\sab)
%   \right|
%   : \theta \in [0,\pi], \sab \in [0,2\pi)
%   \right\} \\
% %   &=
% %   {\rm max} \left\{
% %   \left|
% %   f(\saa_{1/\scale},\sab)
% %   \right|
% %   : \theta \in [0,\pi], \sab \in [0,2\pi)
% %   \right\} \\
%   &= \| f \|_\infty
%   \spcend .
%   \end{split}
%   \end{displaymath}
% \fi
% Thus \eqn{\ref{eqn:cocycle}} gives the correct form for the 
% \pnormtext\ preserving cocycle for all positive integer values 
% of \pnorm.%, including $\pnorm=\infty$.

\section{Proof of MF Theorem (Theorem 1)}
\label{sec:appn_mf}

We solve the \mf\ optimisation problem by minimising the Lagrangian 
\begin{align*}
\lagrn &= 
\displaystyle
\sumlmb \:
\noisecl_\el  \:
\left | \shc{(\fil_\scalepnorm)}{\el}{m} \right |^2  
+ 
\lagrnmult_1 \,
\zreal \!
\left [
\sumlmb
\shc{\tmpl}{\el}{m} \:
\shc{(\fil_\scalepnorm)}{\el}{m}^\cconj - 1
\right ] \\
& \quad
\displaystyle
\mbox{} +
\lagrnmult_2 \,
\zimag \!
\left [
\sumlmb
\shc{\tmpl}{\el}{m} \:
\shc{(\fil_\scalepnorm)}{\el}{m}^\cconj
\right ]
\spcend ,
\end{align*}
where $\lagrnmult_1$ and $\lagrnmult_2$ are Lagrange multipliers. 
Notice that the real and imaginary parts of the constraint
are made explicit.  To solve this problem the filter and template 
spherical harmonic coefficients are represented in terms of their real and imaginary
parts: let 
\mbox{$\shc{(\fil_\scalepnorm)}{\el}{m}=\za{\el}{m}+\img\zb{\el}{m}$} and
\mbox{$\shc{\tmpl}{\el}{m}=\zc{\el}{m}+\img\zd{\el}{m}$}.
% The Lagrangian is then given by
% \begin{align*}
% \lagrn &= 
% \displaystyle
% \sum_{\el =0}^{\infty} \: \sum_{m=-\el }^{\el } \:
% \noisecl_\el  \:
% (\za{\el}{m}^2 + \zb{\el}{m}^2) \\
% & \quad + 
% \lagrnmult_1 \,
% \left [
% \sum_{\el =0}^{\infty} \:
% \sum_{m=-\el }^{\el } \:
% (\za{\el}{m} \zc{\el}{m} + \zb{\el}{m} \zd{\el}{m}) - 1
% \right ] \\
% & \quad
% \displaystyle
% \mbox{} +
% \lagrnmult_2 \,
% \sum_{\el =0}^{\infty} \:
% \sum_{m=-\el }^{\el } \:
% (\za{\el}{m} \zd{\el}{m} - \zb{\el}{m} \zc{\el}{m})
% \spcend .
% \end{align*}
To minimise the Lagrangian one requires
\begin{displaymath}
\label{eqn:mfoptf}
\frac{\partial \lagrn}{\partial \za{\el}{m}}
=
2 \noisecl_{\el} \, \za{\el}{m}
+ \lagrnmult_1 \zc{\el}{m}
+ \lagrnmult_2 \zd{\el}{m}
= 0
\end{displaymath} 
and 
\begin{displaymath}
\frac{\partial \lagrn}{\partial \zb{\el}{m}}
=
2 \noisecl_{\el} \, \zb{\el}{m}
+ \lagrnmult_1 \zd{\el}{m}
- \lagrnmult_2 \zc{\el}{m}
= 0
\spcend ,
\end{displaymath} 
plus the original constraints specified by \eqn{\ref{eqn:mf_constraint}} (one for the real and one for the imaginary component).
% \begin{displaymath}
% \frac{\partial \lagrn}{\partial \lagrnmult_1}
% =
% \sum_{\el =0}^{\infty} \:
% \sum_{m=-\el }^{\el } \:
% (\za{\el}{m} \zc{\el}{m} + \zb{\el}{m} \zd{\el}{m}) - 1
% = 0
% \end{displaymath}
% and 
% \begin{displaymath}
% \label{eqn:mfoptl}
% \frac{\partial \lagrn}{\partial \lagrnmult_2}
% =
% \sum_{\el =0}^{\infty} \:
% \sum_{m=-\el }^{\el } \:
% (\za{\el}{m} \zd{\el}{m} - \zb{\el}{m} \zc{\el}{m})
% = 0
% \spcend .
% \end{displaymath}
Solving these equations simultaneously, one finds
\mbox{$\lagrnmult_1 = -2\filvara^{-1}$} and $\lagrnmult_2 = 0$
for the Lagrange multipliers and 
$\za{\el}{m} = \filvara^{-1} \noisecl_\el^{-1} \zc{\el}{m}$
and 
$\zb{\el}{m} = \filvara^{-1} \noisecl_\el^{-1} \zd{\el}{m}$
for the real and imaginary parts of the filter spherical harmonic
coefficients, where $\filvara$ is defined by \eqn{\ref{eqn:filvara}}.
Thus, the spherical harmonic coefficients of the optimal \mf\ on the
sphere are given by 
$
\shc{(\fil_\scalepnorm)}{\el}{m}=
\filvara^{-1} \noisecl_\el^{-1}
\shc{\tmpl}{\el}{m}
$.

The extremum found is checked to ensure it is a minimum.
The second derivatives of the Lagrangian are 
$\lagrn_{aa} = 2 \noisecl_\el$,
$\lagrn_{bb} = 2 \noisecl_\el$ and
$\lagrn_{ab} = 0$, where the subscript notation for partial
derivatives is used. Now  
$\lagrn_{ab}^2 < \lagrn_{aa} \lagrn_{bb}$ and both \mbox{$\lagrn_{aa}>0$} and 
\mbox{$\lagrn_{bb}>0$}, hence a minimum has indeed been found.

\section{Proof of SAF Theorem (Theorem 2)}
\label{sec:appn_saf}

We solve the \saf\ optimisation problem by minimising the Lagrangian 
\begin{align*}
  \lagrn &= 
  \displaystyle
  \sumlmb
  \noisecl_\el  \:
  \left | \shc{(\fil_\scalenautpnorm)}{\el}{m} \right |^2   + 
  \displaystyle
  \lagrnmult_1 \,
  \zreal \!
  \left [
  \sumlmb
  \shc{\tmpl}{\el}{m} \:
  \shc{(\fil_\scalenautpnorm)}{\el}{m}^\cconj - 1
  \right ]  \\
  & \quad
  \displaystyle
  \mbox{} +
  \lagrnmult_2 \,
  \zimag \!
  \Biggl [
  \sumlmb
  \shc{\tmpl}{\el}{m} \:
  \shc{(\fil_\scalenautpnorm)}{\el}{m}^\cconj
  \Biggr ] \nonumber \\
  & \quad \mbox{} + 
  \displaystyle
  \lagrnmult_3 \,
  \zreal \!
  \Biggl [
  \sumlmb
  \shc{(\fil_\scalenautpnorm)}{\el}{m}^\cconj
  (
  \fconsta
  \shc{\tmpl}{\el}{m}  -
  \fconstb
  \shcsp{\tmpl}{\el-1}{m}
  )
  \Biggr ] \\
  & \quad \mbox{} + 
  \displaystyle
  \lagrnmult_4 \,
  \zimag \!
  \Biggl [
  \sumlmb
  \shc{(\fil_\scalenautpnorm)}{\el}{m}^\cconj
  (
  \fconsta
  \shc{\tmpl}{\el}{m}  -
  \fconstb
  \shcsp{\tmpl}{\el-1}{m}
  )
  \Biggr ]
  \spcend ,
  \end{align*}
where $\lagrnmult_1$, $\lagrnmult_2$, $\lagrnmult_3$ and $\lagrnmult_4$ are
Lagrange multipliers.  Notice that the real and imaginary
parts of each constraint are again made explicit.
To solve this problem the filter and template
spherical harmonic coefficients are represented in terms of their real and imaginary
parts: let  \linebreak[5]
\mbox{$\shc{(\fil_\scalenautpnorm)}{\el}{m}=\za{\el}{m}+\img\zb{\el}{m}$} and
\mbox{$\shc{\tmpl}{\el}{m}=\zc{\el}{m}+\img\zd{\el}{m}$}.
To minimise the Lagrangian one requires
\begin{align}
\label{eqn:safld1}
\frac{\partial \lagrn}{\partial \za{\el}{m}} 
&= 
2 \noisecl_{\el} \za{\el}{m}
 + \lagrnmult_1 \zc{\el}{m}
+ \lagrnmult_2 \zd{\el}{m} \nonumber \\
&\quad\mbox{} + \lagrnmult_3 (
     \fconsta\zc{\el}{m} - \fconstb\zcsp{\el-1}{m}
  )  \nonumber \\
&\quad \mbox{} + \lagrnmult_4 (
     \fconsta\zd{\el}{m} - \fconstb\zdsp{\el-1}{m}
  )
= 0 
\end{align}
and
\begin{align}
\label{eqn:safld2}
\frac{\partial \lagrn}{\partial \zb{\el}{m}} 
= &  
2 \noisecl_{\el} \zb{\el}{m}
+ \lagrnmult_1 \zd{\el}{m}
- \lagrnmult_2 \zc{\el}{m} \nonumber\\
& + \lagrnmult_3 (
     \fconsta\zd{\el}{m} - \fconstb\zdsp{\el-1}{m}
  ) \nonumber\\
& \mbox{} - \lagrnmult_4 (
     \fconsta\zc{\el}{m} - \fconstb\zcsp{\el-1}{m}
  )
= 0
\spcend ,
\end{align}
plus the original constraints specified by \eqn{\ref{eqn:saf_constraint_i}} and \eqn{\ref{eqn:saf_constraint_ii}}.  Using \eqn{\ref{eqn:safld1}} and
\eqn{\ref{eqn:safld2}} it is possible to eliminate  $\za{\el}{m}$ and  $\zb{\el}{m}$
from the original constraints, yielding a system of linear equations for
the Lagrange multipliers that depend on the template spherical
harmonic coefficients only.
% \begin{displaymath}
% \left [ 
% \begin{array}{cccc}
%  \filvara & 0 & \zreal(\filvarb) & -\zimag(\filvarb) \\
%  0 & \filvara & \zimag(\filvarb) & \zreal(\filvarb) \\
%  -\zreal(\filvarb) & -\zimag(\filvarb) & -\filvarc & 0 \\
%  \zimag(\filvarb) & -\zreal(\filvarb) & 0 & -\filvarc 
% \end{array}
% \right ]\!\!\!
% \left [ 
% \begin{array}{c}
%  \lagrnmult_1 \\
%  \lagrnmult_2 \\
%  \lagrnmult_3 \\
%  \lagrnmult_4
% \end{array}
% \right ]
% =
% \left [ 
% \begin{array}{c}
%  -2 \\
%  0 \\
%  0 \\
%  0
% \end{array}
% \right ] \!\!\! ,
% \end{displaymath}
% where the filter variables $\filvara$, $\filvarb$ and $\filvarc$ are
% defined by  \eqn{\ref{eqn:filvara}}, \eqn{\ref{eqn:filvarb}} and 
% \eqn{\ref{eqn:filvarc}} respectively.
%
Solving this system one finds
 $[\lagrnmult_1,
 \lagrnmult_2,
 \lagrnmult_3,
 \lagrnmult_4] = 
 \frac{2}{\filvardenom}
 [
 -\filvarc,
 0,
 \zreal(\filvarb),
 -\zimag(\filvarb)  
 ]$
% \begin{displaymath}
% \left [ 
% \begin{array}{c}
%  \lagrnmult_1 \\
%  \lagrnmult_2 \\
%  \lagrnmult_3 \\
%  \lagrnmult_4
% \end{array}
% \right ]
% =
% \frac{2}{\filvardenom}
% \left [ 
% \begin{array}{c}
%  -\filvarc \\
%  0 \\
%  \zreal(\filvarb) \\
%  -\zimag(\filvarb) \\
% \end{array}
% \right ]
% \end{displaymath}
for the Lagrange multipliers.
% , where the filter variables $\filvara$, $\filvarb$, $\filvarc$ and $\filvardenom$
% are defined by  \eqn{\ref{eqn:filvara}}, \eqn{\ref{eqn:filvarb}}, \eqn{\ref{eqn:filvarc}} and  \eqn{\ref{eqn:filvardenom}} respectively.  
Substituting the Lagrange multipliers
into \eqn{\ref{eqn:safld1}} and \eqn{\ref{eqn:safld2}} one obtains the
following expressions for the real and imaginary parts of the filter
spherical harmonic coefficients:
\begin{align*}
\za{\el}{m} 
= 
(\filvardenom \noisecl_\el )^{-1}
\bigl [ &
\filvarc \:
\zc{\el}{m}   - \zreal(\filvarb) 
( \fconsta\zc{\el}{m} - \fconstb\zcsp{\el-1}{m} 
)  \\
&  \mbox{} + \zimag(\filvarb) 
( \fconsta\zd{\el}{m} - \fconstb\zdsp{\el-1}{m} 
)
\bigr ]
\end{align*}
and 
\begin{align*}
\zb{\el}{m}
= 
(\filvardenom \noisecl_\el)^{-1}
\bigl[ &
\filvarc \:
\zd{\el}{m} 
- \zreal(\filvarb) 
( \fconsta\zd{\el}{m} - \fconstb\zdsp{\el-1}{m} 
) 
\\
&
- \zimag(\filvarb) 
( \fconsta\zc{\el}{m} - \fconstb\zcsp{\el-1}{m} 
)
\bigr]
\spcend .
\end{align*}
Thus, combining the real and imaginary parts, the spherical harmonic
coefficients of the optimal \saf\ are given by \eqn{\ref{eqn:saf}}.

The extremum found is checked to ensure it is a minimum.
The second derivatives of the Lagrangian are 
$\lagrn_{aa} = 2 \noisecl_\el$,
$\lagrn_{bb} = 2 \noisecl_\el$ and
$\lagrn_{ab} = 0$, where the subscript notation for partial
derivatives is used. Now  
$\lagrn_{ab}^2 < \lagrn_{aa} \lagrn_{bb}$ and both \mbox{$\lagrn_{aa}>0$} and 
\mbox{$\lagrn_{bb}>0$}, hence a minimum has indeed been found.

\bibliographystyle{IEEEtran}
\bibliography{bib}
%\begin{thebibliography}{1}

%H.~Kopka and P.~W. Daly, \emph{A Guide to {\LaTeX}}, 3rd~ed.\hskip 1em plus

%\end{thebibliography}

%\begin{biography}[{\includegraphics[width=1in,height=1.25in,clip,keepaspectratio]{mshell}}]{Michael Shell}
% where an .eps filename suffix will be assumed under latex, and a .pdf suffix
% will be assumed for pdflatex; or if you just want to reserve a space for
% a photo:

\vspace*{-10mm}

\begin{biography}[{\includegraphics[width=1in,height=1.25in,clip,keepaspectratio]%
{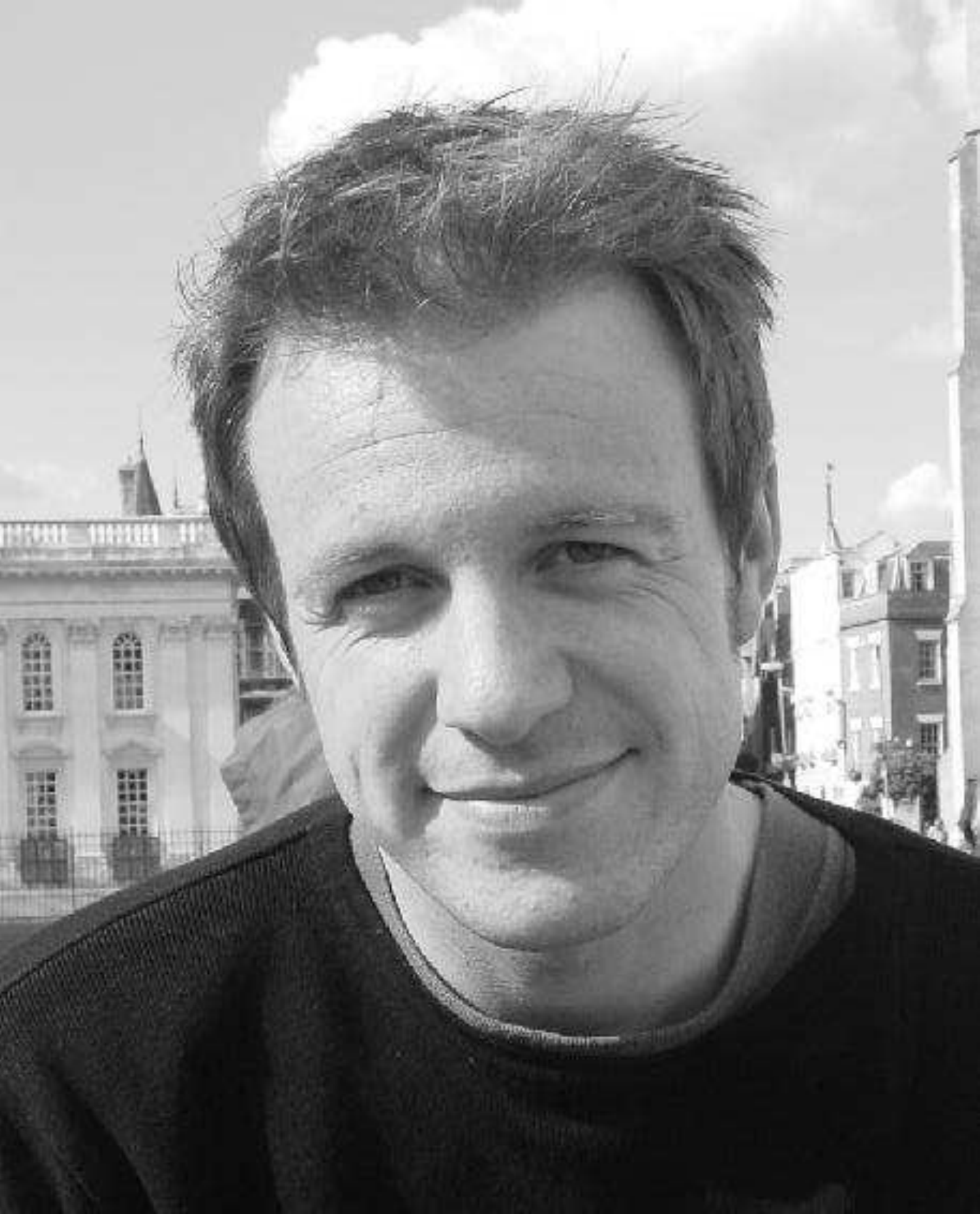}}]{Jason McEwen}
was born in Wellington, New Zealand, in August 1979.
He received a B.E.\ (Hons) degree in Electrical and Computer Engineering
from the University of Canterbury, New Zealand, in 2002 and a Ph.D.\ degree in Astrophysics from the University of Cambridge in 2006. 

He began a Research Fellowship at Clare College, Cambridge, in
2007. His research interests include wavelets on the sphere, the cosmic
microwave background and wavelet based reflectance and illumination in
computer graphics.
\end{biography}

\vspace*{-10mm}

\begin{biography}[{\includegraphics[width=1in,height=1.25in,clip,keepaspectratio]%
{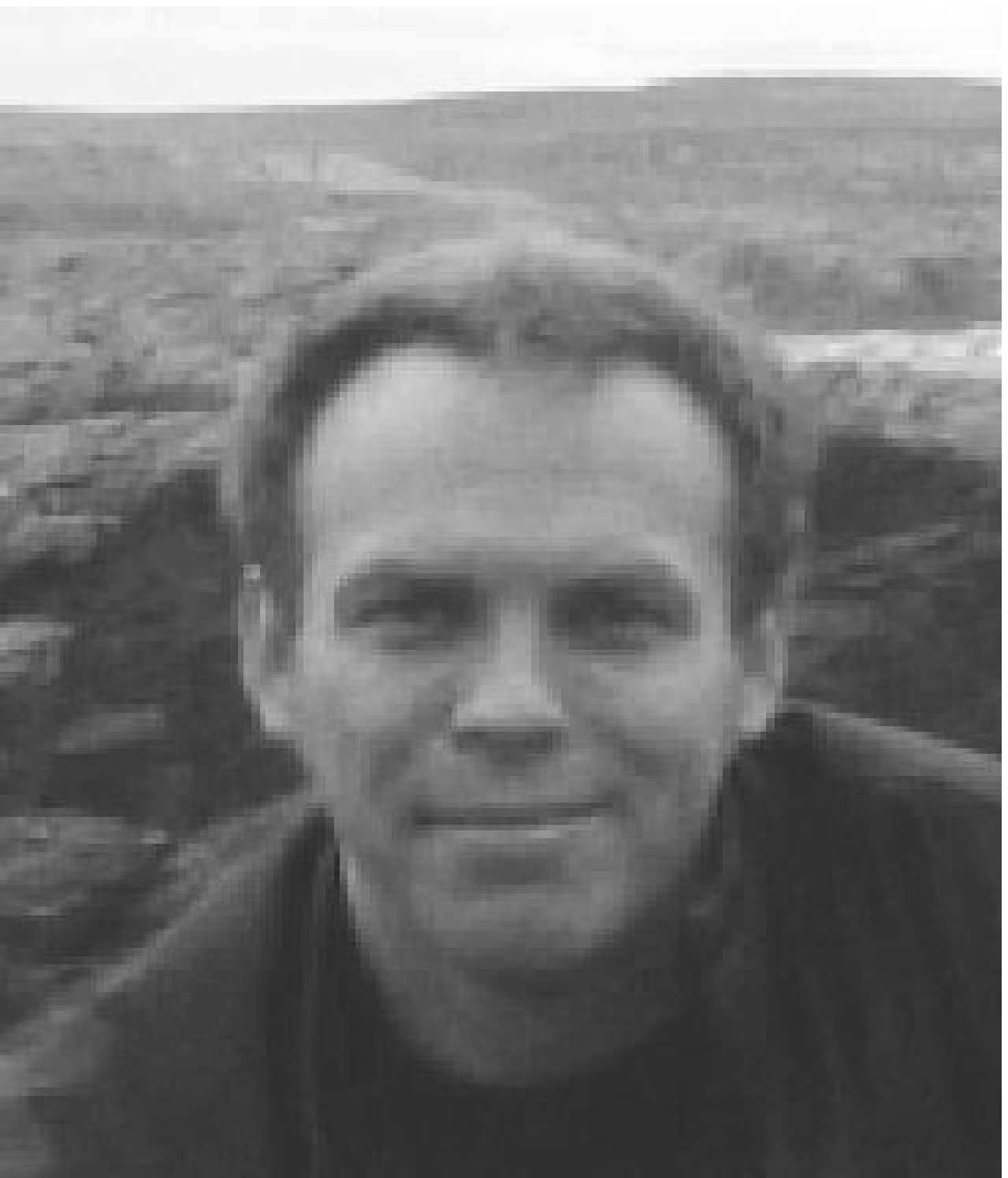}}]{Michael Hobson} 
was born in Birmingham, England, in September 1967. He
received the B.A.\ degree in Natural Sciences with honours and
the Ph.D.\ degree in Astrophysics from the University of
Cambridge, England, in 1989 and 1993 respectively.

Since 1993, he has been a member of the Astrophysics Group of the
Cavendish Laboratory at the University of Cambridge, where he has been
a Reader in Astrophysics and Cosmology since 2003. His research
interests include theoretical and observational cosmology,
particularly anisotropies in the cosmic microwave background,
gravitation, Bayesian analysis techniques and theoretical optics. 
\end{biography}

% insert where needed to balance the two columns on the last page
%\newpage
\vspace*{-10mm}

\begin{biography}[{\includegraphics[width=1in,height=1.25in,clip,keepaspectratio]%
{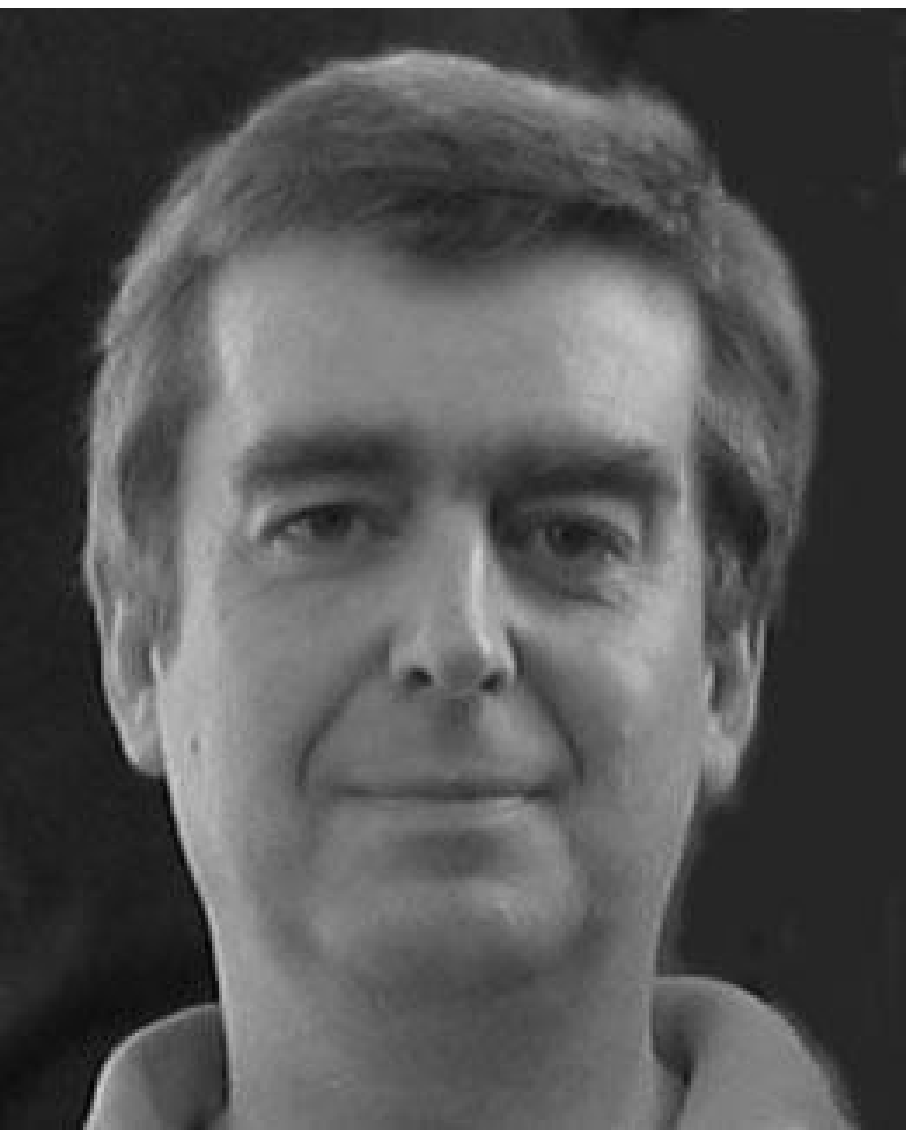}}]{Anthony Lasenby}
was born in Malvern, England, in June 1954. He
received a B.A.\ then M.A.\ from the University of Cambridge in
1975 and 1979, an M.Sc.\ from Queen Mary College, London in 1978
and a Ph.D.\ from the University of Manchester in 1981. 

His Ph.D.\
work was carried out at the Jodrell Bank Radio Observatory
specializing in the Cosmic Microwave Background, which has been a
major subject of his research ever since. After a brief period at
the National Radio Astronomy Observatory, Tucson, Arizona, he moved
from Manchester to Cambridge in 1984, and has been at the
Cavendish Laboratory Cambridge since then. He is currently Head
of the Astrophysics Group and the Mullard Radio Astronomy
Observatory  of the Cavendish Laboratory.  His other main interests include theoretical physics
and cosmology, the application of new geometric techniques in
computer graphics and electromagnetic modelling, and statistical
techniques  in data analysis.
\end{biography}


%\vfill


\end{document}